\documentstyle [aps,pre,epsf]{revtex} 

\newcommand{\rr  }   { {\bf r} }
\newcommand{\dr  }   { {\rm d}\rr\ }
\newcommand{\dx  }   { {\rm d}x\ }
\newcommand{\e   }   { {\rm e} }

\begin{document}
\setlength{\baselineskip}{18pt}
\draft

%-----------------------------------------------------------
% 1st page
%-----------------------------------------------------------
\title {\bf The Effect of Polyelectrolyte Adsorption on 
Inter--Colloidal Forces}

\author {Itamar Borukhov and David Andelman$^{*}$}
\address{School of Physics and Astronomy, \\
         Raymond and Beverly Sackler
         Faculty of Exact Sciences, \\
         Tel-Aviv University,
         Ramat-Aviv 69978, Tel-Aviv, Israel}

\author {Henri Orland}
\address{Service de Physique Th\' eorique, \\
         CE-Saclay,
         91191 Gif-sur-Yvette Cedex, France }

\date{\today}

\maketitle

%- - - - - - - - - - - - - - - - - - - - - - - - - - - - - -

\begin{abstract}
\setlength{\baselineskip}{12pt}
   The behavior of polyelectrolytes between charged surfaces
immersed in semi--dilute solutions is investigated theoretically.
   A continuum mean field approach is used for 
calculating numerically concentration profiles 
between two electrodes held at a constant potential.
   A generalized contact theorem relates the 
inter--surface forces to the concentration profiles.
The numerical results show that over-compensation of 
the surface charges by adsorbing polyelectrolytes 
can lead to effective {\it attraction} between 
{\it equally charged} surfaces.
    Simple scaling arguments enable us to characterize qualitatively 
the inter--surface interactions as function of the fraction
of charged monomers $p$ and the salt concentration $c_b$.
In the low salt regime 
we find strong repulsion at short distances, 
where the polymers are depleted from the inter--surface gap, 
followed by strong attraction when the two adsorbed layers overlap.
The magnitude of this attraction scales as $p^{1/2}$
and its dominant length scale is proportional to $ a/p^{1/2}$, 
where $a$ is the monomer size.
At larger distances the two adsorbing surfaces interact 
via a weak electrostatic repulsion.
For strong polyelectrolytes at high salt concentration 
the polymer contribution to attraction at short distances
scales as $p/c_b^{1/2}$ and the length scale is
proportional to $\kappa_s a^2/p$, where 
 $\kappa_s^{-1}$ is the Debye--H\"uckel screening length.
For weak polyelectrolytes at high salt concentration 
the interaction is repulsive for all surface separations 
and decays exponentially with a decay length equal to 
 $\kappa_s^{-1}$.
The effect of irreversible adsorption is discussed as well and
it is shown that inter--surface attraction can be obtained 
in this case as well.
\end{abstract}
%- - - - - - - - - - - - - - - - - - - - - - - - - - - - - -
%  PACS. 61.25H&--&Macromolecular and polymer solutions; 
%                  polymer melts; \\
%              &  &Swelling. \\
%  PACS. 68.10 &--&Fluid surfaces and interfaces with fluids
%or: 
%  PACS. 36.20 &--&Macromolecules and polymer molecules
%  PACS. 41.10D&--&Electrostatics, Magnetostatics \\

%-----------------------------------------------------------
% 1st section
%-----------------------------------------------------------
% \pagebreak
\setlength{\baselineskip}{18pt}
\pagestyle{plain}
%-----------------------------------------------------------
\section{Introduction}
%-----------------------------------------------------------

   Polymers are known to affect the interactions
of colloidal particles  in solution \cite{Joanny,Cabane,Dickinson}.
Adsorption of charged polymers ({\em polyelectrolytes}) 
to oppositely charged colloids may turn inter--colloidal 
repulsion into attraction, leading to flocculation.
This phenomenon is used in industrial applications such as 
water filtration, paper making and mineral processing.
   The reversed process is useful as well, since adsorbed 
polyelectrolytes (in different conditions) can also 
stabilize colloidal suspensions such as 
paint, ink or medical suspensions against attractive forces 
(e.g., van der Waals forces).

%- - - - - - - - - - - - - - - - - - - - - - - - - - - - - -

   One of the most common
techniques to study experimentally 
the adsorption of polyelectrolytes between two  
surfaces is the
{\em Surface Force Apparatus} (SFA) \cite{Israelachvili},
which allows delicate measurements of inter-surface forces 
at distances as small as a few Angstroms.
   In these experiments 
\cite{Klein,Marra,Claesson,Dahlgren-etal,Audebert,Eriksson,Dahlgren,Helm}, 
 attractive and repulsive forces
have been observed, depending on specific details,
such as the type of polyelectrolyte, its concentration, 
the ionic strength of the solution, etc.
   In other experiments it is possible to measure the 
disjoining pressure 
of thin liquid films in the presence of polyelectrolytes and
 as a function of the film  width \cite{Bergeron}. 
Both repulsive and attractive forces have been measured
using this method.

%- - - - - - - - - - - - - - - - - - - - - - - - - - - - - -

  Adsorption of polyelectrolytes was treated theoretically 
in a discrete model 
\cite{vanderSchee,Papenhuijzen,Evers,vandeSteegSim}, 
where the state of the system is described by 
occupation fractions of monomers, ions and solvent 
molecules on a discrete lattice. 
Within mean field, the equilibrium state of the system 
corresponds to the maximal contribution to the 
partition function of the system, and can be calculated
numerically.
 B\"ohmer et al. \cite{Bohmer} have used this model 
to calculate force curves at relatively short distances
(up to 30 molecular layers).
In addition, Monte Carlo computer simulations of polyelectrolytes 
between flat surfaces \cite{JonnsonFlat} and
between charged spheres \cite{JonssonSpheres} 
provide  valuable hints concerning the complex
behavior of polyelectrolytes. However, 
they are limited to relatively short chains and small 
inter-surface distances due to computation time limitations.

Another theoretical approach is a continuum one
\cite{Muthukumar,Varoqui91,Varoqui93,epl,epr,macromols,Podgornik,Chatellier}
where the concentration of the different species are 
taken to be continuous functions of the spatial coordinates.
The mean--field state can be calculated by solving 
two differential equations 
for the polymer concentration and the electrostatic potential
derived using a variational procedure.
Varoqui et al. \cite{Varoqui91,Varoqui93} used the continuum approach 
to investigate polyelectrolyte adsorption onto one surface, 
while Podgornik \cite{Podgornik}
used a similar formalism to calculate inter--surface forces. 
In those works the non-linear excluded volume interaction between the
monomers has not been considered.
Ch\^atellier and Joanny \cite{Chatellier} used a linearized 
version of a similar approach to study the inter--surface 
interactions for polyelectrolytes in a poor solvent.
Recently, we have been able to derive some simple scaling 
relations for the adsorption of polyelectrolytes onto a single charged
surface \cite{macromols}.
These scaling relations were compared to the exact numerical solutions
of the differential equations and to existing
experimental results.
The agreement was reasonable in  two opposite limits: 
(i) low salt concentration (no electrostatic screening) and 
(ii) high salt concentration (strong screening).

   In the present work, the continuum model is used to study 
polyelectrolyte adsorption between two parallel surfaces. 
The advantage of our model is that 
the connectivity of the polymer chains, 
the excluded volume repulsion between monomers in a good solvent
and the Coulomb interactions between the charged monomers, 
counter-ions, co-ions and surface charges
are all taken into account.
The mean field equations are solved numerically in order to
obtain concentration profiles, and the inter--surface interactions
are then calculated from the free energy.
  In addition we extend our earlier scaling approach 
of one adsorbing surface \cite{macromols} to the case of 
two interacting surfaces. The result is a qualitative description 
of the inter--surface interactions as function of
the polyelectrolyte charge and the amount of salt in the solution. 

  The paper is organized as follows:
in the next section we describe the mean--field approach.
In Sec.~III we present numerical results obtained from 
solving the mean field equations,
and in Sec.~IV we use simple scaling arguments to describe the 
inter--surface interactions.
In Sec.~V we study the effect of irreversible adsorption both 
numerically and analytically and
in Sec.~VI we compare our results with experiments.
Finally we present our conclusions and some future prospects.

%-----------------------------------------------------------
\section{The Mean Field Approach}
%-----------------------------------------------------------

%- - - - - - - - - - - - - - - - - - - - - - - - - - - - - -
\subsection{The Basic Equations}
%- - - - - - - - - - - - - - - - - - - - - - - - - - - - - -

  The model system consists of a semi--dilute solution of 
polyelectrolytes in a good solvent placed between two flat 
surfaces (Fig.~1). The solution contains charged 
polymer chains, counter-ions and a monovalent electrolyte (salt).
Having in mind the experimental setup of the surface force 
apparatus (SFA) discussed below, we consider a system 
which is 
coupled to a bulk reservoir of polyelectrolyte chains and salt.
  
  As discussed elsewhere \cite{epl,epr}, the charge distribution 
along the chains depends on the type of polyelectrolyte
as well as on the local conditions, such as the pH and the 
electrostatic potential. 
However, at low electrostatic potentials $|\beta e\psi|\ll 1$,
where $\psi$ is the electrostatic potential, 
 $\beta=1/k_BT$ the inverse thermal energy and $e$ the electron 
charge, the differences between the various charge distribution 
models are small.
We therefore assume hereafter a uniform charge
distribution along the polymer chains with a fractional 
charge $p e$ attached to each monomer.

  In the mean field approach, the free energy of the system 
is expressed in terms of the local electrostatic potential 
 $\psi(\rr)$ at a point $\rr$
and the polymer order parameter $\phi(\rr)$ which is related to 
the local monomer concentration through $c_m(\rr)=|\phi(\rr)|^2$.
The relation between the polymer order parameter and the 
monomer concentration is analogous to the relation between 
the wave function and the probability density 
of a particle in quantum mechanics.
The excess free energy with respect to the bulk can be divided 
into three contributions \cite{Varoqui91,Varoqui93,epl,epr}:
\begin{equation}
   F = \int f(\rr) \dr
     =  \int \biggl\{ f_{\rm pol}(\rr) + f_{\rm ions}(\rr)
                        + f_{\rm el}(\rr) \biggl\} \dr
   \label{F}
\end{equation}
  The polymer contribution is
\begin{equation}
   f_{\rm pol}(\rr) = k_BT \biggl[  \frac{a^2}{6}|\nabla\phi|^2
                 + \frac{1}{2} v (\phi^4-\phi_b^4) \biggr]
                 - \mu_p (\phi^2-\phi_b^2)
        \label{fpol}
\end{equation}
where the first term is the polymer response to local variations 
of the concentration and is due to the connectivity of the 
polymer chain, $a$ being the effective monomer size. 
The second term represents the short ranged monomer--monomer 
interaction and can be viewed as representing an effective volume of 
a single monomer.
For a polymer in a good solvent $v$ is positive. 
However, since $v$ represents an {\em effective}
monomer--monomer interaction it can also be 
negative (in a poor solvent) or 
zero (in a theta solvent) requiring higher order terms 
in $\phi^2$ to be included in the free energy.
For simplicity we will limit ourselves to good solvent conditions
but the formalism can be easily generalized to 
other conditions as well.
The last term couples the system to a reservoir,
 $\mu_p$ being the polymer chemical potential and
 $\phi_b^2$ the bulk monomer concentration.

   The non--electrostatic contribution of the small (monovalent) 
ions is due to their translation entropy and is equal to
\begin{equation}
   f_{\rm ions}(\rr) = \sum_{i=\pm} \biggl\{ k_BT
                   \biggl[ c^i\ln c^i - c^i
                        - \bigr(c^i_b\ln c^i_b - c^i_b\bigr) \biggr]
                 - \mu^i \bigr( c^i-c^i_b \bigr) \biggr\}
        \label{fent}
\end{equation}
   where $c^i(\rr)$ is the local concentration 
of the $i=\pm$ ions (cations and anions)
and $c^i_b$, $\mu^i$ are the bulk concentration and chemical 
potential, respectively.
In the most general case the solution contains two types of
negative ions: the counter-ions which dissociate from the 
polymer chains and the salt anions.
In the reservoir the concentration of negative ions has two 
contributions $c^-_b = c_b+p\phi_b^2$ where $c_b$ is the 
electrolyte bulk concentration, while for the positive ions 
$c^+_b = c_b$.
In principle, one could consider the 
two types of negative ions separately, but for clarity
we take the two types of negative ions to be identical.
%but as far as electrostatic interactions are concerned, 
%separating the two contributions does not affect the final results. 

Finally, the electrostatic contribution is
\begin{equation}
   f_{\rm el}(\rr)  = p e\phi^2\psi +e c^+\psi -e c^-\psi
                 - \frac{\varepsilon}{8\pi}|\nabla \psi|^2
        \label{fel}
\end{equation}
  The first term is the electrostatic energy of charged monomers.
The next two terms represent the positive and negative ions, 
respectively, and the last term is the self energy of the 
electric field where $\varepsilon$ is the dielectric constant 
of the solution.
The sum of the electrostatic contributions can be integrated 
by parts using the Poisson equation (derived below) and yields
 $F_{\rm el}=(\varepsilon/8\pi) \int |\nabla \psi|^2  \dr$, 
as expected, plus electrostatic surface terms. 
%The different terms in eq.~\ref{fel} need to be written explicitly
%in order that the free energy would be a functional 
%of both the electric field (through $\psi$) 
%and the charge densities (through $\phi$ and $c^\pm$). 

   Minimization of the free energy with respect to $c^\pm$, $\phi$
and $\psi$ yields a Boltzmann distribution
for the concentration of the small ions,
 $c^\pm(\rr)=c^\pm_b \exp(\mp\beta e\psi)$, and two
coupled
differential equations for $\phi$ and $\psi$ \cite{epr}:
\begin{equation}
   \nabla^2\psi(\rr)
    = \frac{8\pi e}{\varepsilon}c_b \sinh(\beta e\psi)
    -  \frac{4\pi e}{\varepsilon}
       \left( p\phi^2 - p\phi_b^2 \e^{\beta e\psi} \right)
   \label{PBs}
\end{equation}
\begin{equation}
   \frac{a^2}{6} \nabla^2\phi(\rr)
   = v(\phi^3-\phi_b^2\phi) + p\phi\beta e\psi
   \label{SCFs}
\end{equation}
   Equation~\ref{PBs} is a generalization of the 
Poisson--Boltzmann equation
including the free ions as well as the charged polymers.
The first term represents the salt contribution and
the second term is due to the
charged monomers and their counter-ions.
 Equation~\ref{SCFs} is a generalization of the self--consistent
field equation of neutral polymers \cite{deGennesBook}.
In the bulk the potential and the polymer concentration
have constant bulk values given by
$\psi=0$ and $\phi=\phi_b$, as can be seen in the above equations.

%- - - - - - - - - - - - - - - - - - - - - - - - - - - - - -
\subsection{Two Interacting Surfaces}
%- - - - - - - - - - - - - - - - - - - - - - - - - - - - - -

  The interaction of two charged surfaces in a solution 
containing only
small ions (electrolyte) without  charged polymers
is well established within the framework 
of the Poisson--Boltzmann equation \cite{IsraelachviliBook}.
The electrostatic interaction between two identically
charged  surfaces is found to be repulsive
within this mean-field like theory \cite{beyondPB}. 
  However, the addition of polyelectrolytes to the solution changes
the picture in a subtle way.
Experiments 
\cite{Klein,Marra,Claesson,Dahlgren-etal,Audebert,Eriksson,Dahlgren,Helm,Bergeron}
show that polyelectrolytes reduce this repulsion and 
might even cause  mutual attraction between the two surfaces.

For simplicity, the surfaces are taken as flat,
homogeneous and parallel to each other in order
 that the physical quantities will depend only on the position $x$
between the surfaces (see Fig.~1). 
The effect of the surfaces is introduced through the
boundary conditions on the polymer order parameter $\phi(x)$ 
and the electrostatic potential $\psi(x)$.
  In this work, both surfaces are assumed to be 
kept at the same  constant potential:
\begin{equation}
  \left. \psi \right|_s = \psi_s
  \label{BCpsis}
\end{equation}
and no monomers are adsorbed on the surfaces
\begin{equation}
  \left. \phi \right|_s = 0
  \label{BCphis}
\end{equation}
  Other boundary conditions could have been considered as well.
For example, if a fixed surface charge $\sigma$ is assumed then the 
electrostatic boundary condition would include the electric field:
 $\left.\psi' \right|_s = - {4\pi \sigma}/{\varepsilon}$.
In real systems neither the surface potential nor the 
surface charge are fixed. The choice of one or another is only 
an approximation whose quality depends on the details of the 
experimental system.
Similarly, for the polymer boundary conditions one could consider
an adsorbing surface instead of a non-adsorbing one 
\cite{deGennesCahn}.

Given these boundary conditions, 
the Poisson--Boltzmann and self-consistent field 
equations (\ref{PBs},\ref{SCFs}) uniquely determine
 $\psi(x)$ and $\phi(x)$.
However, experiments usually probe global properties such as 
the amount of monomers adsorbed per unit area or the 
inter--surface interactions.

   The total amount of monomers (per unit area) 
between the two surfaces $\Gamma(w)$ can be easily 
calculated from the polymer concentration profile since 
\begin{equation}
  \label{Gamma}
  \Gamma(w) = \int_{-w/2}^{w/2} \phi^2(x) \dx
\end{equation}
Another measure for the strength of the adsorption is the 
average monomer concentration divided by the bulk concentration:
\begin{equation}
  \label{u2}
  \left< {\phi^2\over\phi_b^2} \right> = 
  {1\over w}\int_{-w/2}^{w/2} {\phi^2(x)\over\phi_b^2} \dx = 
  {\Gamma(w) \over w\phi_b^2}
\end{equation}
The latter quantity relates to  the 
strength of the adsorption only at small distances. 
As demonstrated by the numerical examples below,  
at larger distances, 
$\Gamma$ saturates to a constant 
and the average concentration decreases as $1/w$.

   In addition, it is of interest to calculate the total amount of 
charge (per unit area)
carried by the adsorbed polymers $\sigma_p(w)=pe\Gamma(w)$
as compared to the induced surface charge density $\sigma_s(w)$
(on a single surface).
The latter depends on the inter-surface distance as we 
have chosen to work with constant surface potentials rather than 
constant surface charges.

   The adsorption of polyelectrolytes strongly affects the 
inter--surface interactions. 
The excess free energy per unit area for two surfaces
at a distance $w$ apart can be calculated 
from the concentration profiles $\phi(x)$ and $\psi(x)$:
\begin{equation}
  \Delta F(w) = \int_{-w/2}^{w/2} f[\phi(x),\psi(x)] \dx 
              -2 F_1
  \label{FreeEnergy}
\end{equation}
where $f(x)$ was introduced in eqs.~\ref{F}-\ref{fel}
and $F(w \to \infty)=2F_1$ is the free energy of two isolated surfaces
at infinite separation.

 The variation of this free energy with respect 
to the inter-surface distance $w$ gives the inter-surface pressure
(or force per unit area):
\begin{equation}
  \Pi(w) = -\frac{\delta(\Delta F)}{\delta w}
  \label{force}
\end{equation}
  It can be shown from eqs.~\ref{FreeEnergy},\ref{force} and from
 $\delta(\Delta F)/\delta\phi(x)=\delta(\Delta F)/\delta\psi(x)=0$
that
\begin{equation}
  \Pi(w) = -f(x=0)
  \label{force0}
\end{equation} 
  where $f(x=0)$ is the free energy density (per unit volume) 
at the mid--plane.
   This relation is a generalization of the contact theorem
of neutral polymers \cite{deGennesForce}. 
In our case the calculation of $f(x=0)$ yields:
\begin{equation}
  \beta\Pi(w) = -p\phi^2(0)y(0)-{1\over 2} v( \phi^2(0)-\phi_b^2)^2
  +p\phi_b^2[\e^{y(0)}-1]+2c_b [\cosh(y(0))-1]
  \label{force1}
\end{equation} 
where $y(0)=\beta e\psi(0)$ is the reduced electrostatic potential at 
the symmetry plane ($x=0$).
The above  expression  is obtained by 
inserting
 the equilibrium 
(Boltzmann) distribution of the small ions 
$c^\pm=c_b^\pm\exp(\mp\beta e \psi)$ back
into the free energy, eqs.~\ref{fpol}-\ref{fel}.

  Equations~\ref{force} and 
\ref{force0} for the force are valid 
for the planar geometry (Fig.~1). 
In some experiments \cite{Bergeron} where the disjoining pressure 
of thin liquid films is measured, the two surfaces are indeed
parallel to each other and $\Pi(w)$ is measured directly.
However, most experiments 
 \cite{Klein,Marra,Claesson,Dahlgren-etal,Audebert,Eriksson,Dahlgren,Helm}
use the surface force apparatus \cite{Israelachvili} where
the force $\Pi_R$ is measured between 
two cylindrical surfaces of radii $R$ with a $90^\circ$ 
tilt between their major axes (see Fig.~2a).
At small distances $w\ll R$ the Derjaguin approximation
\cite{IsraelachviliBook}
relates the measured force to the {\em excess free energy}
(as given by eq.~\ref{FreeEnergy}) and {\em not} to its 
derivative (as given by eqs.~\ref{force},\ref{force0}) 
\begin{equation}
  \label{SFA}
  {\Pi_R(w)\over R} = 2\pi \Delta F(w)
\end{equation}
For clarity purposes we denote 
the force 
per unit area acting between two infinite flat surfaces
as $\Pi(w)$ (eqs.~\ref{force},\ref{force0}),
and the absolute force acting between 
two cross cylinders as $\Pi_R(w)$.

The Derjaguin approximation can also be used to 
calculate the interaction between two spheres of radii $R$ 
at small distances $w\ll R$ (Fig.~2b)
as is the case for colloidal suspensions.
The inter--colloidal force is related to the interaction 
free energy of two flat surfaces but with a different numerical 
factor:
\begin{equation}
  \label{spheres}
  {\Pi_R(w) \over R} = \pi \Delta F(w)
\end{equation}

%-----------------------------------------------------------
\section{Numerical Results}
%-----------------------------------------------------------
%- - - - - - - - - - - - - - - - - - - - - - - - - - - - - -
\subsection{Low Salt Concentration}
%- - - - - - - - - - - - - - - - - - - - - - - - - - - - - -

%- - - - - - - - - - - - - - - - - - - - - - - - - - - - - -
\noindent{\bf Concentration Profiles:}~~~
  In order to solve the mean--field equations 
(eqs.~\ref{PBs},\ref{SCFs}), we use a minimal squares scheme 
in which the spacing between the two surfaces is divided into 
 $N\sim 100$ intervals. An error functional which sums the squares
of the local errors in eqs.~\ref{PBs},\ref{SCFs} is minimized 
with respect to the values of $\phi$ and $\psi$ at the discrete 
grid points, under the constraint of the boundary conditions 
eqs.~\ref{BCpsis},\ref{BCphis}.

  Typical solutions are presented in Fig.~3.
The polymer is positively charged ($p=1$) and attracted to
non-adsorbing surfaces held at a constant negative potential
 ($\psi_s<0$). Here we focus on the weakly screened limit 
(low salinity).
The effect of screening can be estimated by comparing the 
inter--surface separation $w$ with the 
 Debye--H\"uckel screening length $\kappa_s^{-1}$ defined 
by $\kappa_s^2=8\pi l_B c_b$ where $l_B=e^2/\varepsilon k_BT$ is the 
Bjerrum length equal to $7$\AA\ for aqueous solutions at 
room temperature.
At low salt concentration, $\kappa_s^{-1}\gg w/2$, screening 
is weak and plays only a minor role, 
whereas  
at high salt concentration, $\kappa_s^{-1}\ll w/2$, screening 
strongly reduces the Coulomb interactions in 
the adsorbed layer.
In Fig.~3, the solution contains a small amount of monovalent salt
($c_b=1$mM) so that the electrostatic screening length 
 $\kappa_s^{-1} \simeq 100$\AA\  is larger than the 
inter--surface distance which varies between 7 and 40\AA.

In Fig.~3a the reduced electrostatic potential $y(x)=\beta e\psi(x)$ 
is plotted as a function of the position $x$ between the two surfaces.
The reduced monomer concentration $\phi^2(x)/\phi_b^2$ is shown
in Fig.~3b.
Despite the fact that the surface potential is not very high,
 $y_s=-2.0$ corresponding to $\psi_s\simeq -50$mV, 
the adsorption is quite strong and the concentration in the 
gap between the two surfaces can increase by three orders 
of magnitude above the bulk concentration.  
The adsorption here is purely electrostatic
since the only source of attraction is
due to the  electrostatic boundary conditions. 
A neutral polymer in similar conditions will not 
adsorb to the surfaces.

  At small inter--surface distances the adsorbed polymers consist 
of a single layer extending from one surface to the other and 
the potential is negative everywhere in the gap. 
As the surfaces are drawn  away from each other,
first the amount of adsorbed polymer grows rapidly
and then the 
adsorbed layer separates into two distinct layers near the two 
surfaces. In addition, the potential changes sign and becomes positive 
in the central region. By integrating the 
Poisson equation from the surface
to the point where the potential changes sign,
it can be easily shown that this
sign reversal is due to an
over--compensation of the surface charges by the 
layer of adsorbed monomers.

The over--compensation is specific to charged polymers and does
not appear in the Poisson-Boltzmann formalism for small ions 
(regular electrolytes). Physically,
charged monomers which adsorb close to the surface 
are connected to other monomers which reside at some 
larger distance. In our model, the polymer chains
resist fluctuations on length
scales smaller than the Edwards correlation length 
 $\xi_E\sim a/\sqrt{vc_m}$ of neutral polymers and thus
over--compensate the surface charges
and cause the potential to reverse sign.
This over-compensation is much more pronounced
when salt is added to the solution as will
be discussed below. In the central region where
the sign of the potential is opposite to that of the surface potential,
the concentration of negative ions is larger than that of 
positive ions. We stress that
when all 
contributions to the charge density are considered 
(polymer and small ions), 
the charges in the solution exactly balance the surface charges
(see Fig.~5).

   At yet larger distances ($w>20$\AA\ for the physical parameters 
of Fig.~3) the two adsorbed layers do not change any more.
This occurs when the inter--surface distance $w$ is larger than 
twice the width of the adsorbed layers. 
The two surfaces are almost decoupled
and
single surface adsorption is recovered.
The polymer concentration between the two adsorbed layers is 
small and comparable to the bulk concentration.
As long as the screening length $\kappa_s^{-1}$ is larger than the 
distance $w$, the electrostatic potential is nearly a constant 
(e.g. $y\simeq 0.2$ in Fig.~3). 
At even larger distances, $w>\kappa_s^{-1}$, 
the effect of screening will show up,
and the mid--plane potential will gradually decay to zero. 

%- - - - - - - - - - - - - - - - - - - - - - - - - - - - - -
\noindent{\bf Polyelectrolyte Adsorption:}~~~
  In Fig.~4a the total amount of monomers (per unit area)
adsorbed between 
the two surfaces $\Gamma(w)$ is plotted as a function of the 
inter-surface distance $w$ for two charge fractions $p=1$
(solid curve) and $p=0.2$ (dashed curve). Similarly,
in Fig.~4b the average reduced monomer concentration 
 $\langle\phi^2/\phi_b^2\rangle$ is plotted as a function of $w$. 
Three regimes can be distinguished in accord with the 
findings presented on Figs.~3-4:
at very short distances ($w\simeq 5$\AA) the confinement 
of the polymer to a narrow slit competes with the electrostatic 
attraction of the charged monomers to the surface and avoids 
strong adsorption. The polymer is not expelled totally from 
the gap between the two surfaces but its concentration 
is of the order of the bulk concentration.

   As the surfaces are taken further apart, the adsorption increases
rapidly until it reaches its maximal value.
At this point the average concentration can be three orders 
of magnitude higher than the bulk concentration.
At larger distances the two surfaces decouple from each other
and
the adsorbed amount decreases 
towards a saturation value. At this stage
the system can be described as  two independent 
layers adsorbing onto the two surfaces. 
The saturation value of $\Gamma$
is approximately twice the adsorbed amount to a 
single surface $\Gamma(w\rightarrow\infty)=2\Gamma_1$.
In a preceding work \cite{macromols} we 
investigated adsorption onto a single surface and
have shown that in 
the low salt limit $\Gamma_1\sim 1/\sqrt{p}$.
This behavior is a result of two competing interactions: 
(i) the electrostatic attraction of the charged monomers 
to the surface which is proportional to $p$; 
and (ii) the Coulomb repulsion between charged monomers in the 
adsorbed layer which is proportional to $p^2$.
For strong polyelectrolytes the latter interaction dominates 
and 
the adsorbed amount $\Gamma_1$ increases as the 
fractional charge $p$ decreases.
This scaling
 behavior is in accord with the saturated values of Fig.~4a.

  In Fig.~5 the charge densities per unit area are plotted 
as functions of the distance $w$ for the same sets of values 
as in Fig.~4. 
Using the single surface results we verify that indeed in the 
saturated  regime of large inter-surface
separations $\sigma_p(\infty) \sim p\Gamma_1 \sim\sqrt{p}$.
Another observation which can be made
from Fig.~5 is that the two charge densities almost balance
each other. 
In the low salt limit these charge densities must cancel 
each other since the amount of salt is too small to 
play any significant role in neutralizing the solution.

%- - - - - - - - - - - - - - - - - - - - - - - - - - - - - -
\noindent{\bf Free Energies and Forces:}~~~
In Fig.~6, inter-surface force profiles were calculated  
for $c_b=10^{-6}$M, corresponding to $\kappa_s^{-1}\simeq 3000$\AA,
and for two values of the polymer charge fraction $p$.
 In Fig.~6a, the excess free energy per unit area $2\pi\Delta F(w)$,
eq.~\ref{FreeEnergy},
which is the physical quantity measured in SFA experiments,
is plotted as a function of the inter--surface distance $w$.
In Fig.~6b the force per unit area $\Pi(w)$, eq.~\ref{force},
acting between flat surfaces is plotted as a function 
of the distance.
This force can be measured directly in disjoining pressure 
experiments of thin films \cite{Bergeron}.

The two surfaces 
strongly repel each other at very short distances and
attract  at short distances.
Note that for small ions such attraction is not present within the 
Poisson-Boltzmann approximation. For polymer chains the lack 
of translational entropy as well as the large correlation
length $\xi_E$ enhance the effective attraction between the 
surfaces at short separations.
As the charge fraction is lowered, the attraction becomes 
weaker and the length scale of attraction increases. 
These two effects can be explained 
by simple arguments which are presented in section IV.
Attractive interactions have been observed experimentally 
\cite{Dahlgren-etal} and were attributed to ``bridging'' of chains
between the surfaces ---
a mechanism which is not present in our approach. 

A secondary repulsion appears at large distances but
is too weak to be shown on a linear scale in our plots. However,
this secondary repulsion can be made quite pronounced, in particular
when the polymer surface excess is fixed at a large value. This
is further discussed in Sec. V. For small ions, the origin of the repulsion
is entropic whereas here it is due to over-compensation of the surface
charges.
%This secondary repulsion can be attributed to Coulomb 
%repulsion of two adsorbed layers on the two surfaces 
%at distances larger than twice the adsorption length.
%The attraction at short distances can be intuitively
%understood in terms of an overlap 
%of the two adsorbed layers. 
%At these distances polymer chains crossing the
%mid-plane contribute an attractive spring--like attraction
%between the two surfaces. 

% A special feature of Fig.~6 is the steep decrease in 
%the pressure $\Pi$ at very short distances.
%The fact that this sharp reduction of the force is accompanied
%by the depletion of polymer from within the gap (Fig.~4)
%indicates that the polymer dominates the inter--surface interactions.

%- - - - - - - - - - - - - - - - - - - - - - - - - - - - - -
\subsection{High Salt Concentration}
%- - - - - - - - - - - - - - - - - - - - - - - - - - - - - -

%- - - - - - - - - - - - - - - - - - - - - - - - - - - - - -
\noindent{\bf Concentration Profiles:}~~~
At high salt concentration the screening length 
$\kappa_s^{-1}$ is smaller 
than the 
inter--surface separation $w$. 
The effect of screening on the concentration profiles is 
demonstrated in Figs.~7 and 8, where a strong polyelectrolyte 
($p=1$) is adsorbed from a solution containing large amounts
of salt, 
 $c_b=1$M for which the screening length is $\kappa_s^{-1}=3$\AA.
In Fig.~7 the reduced electrostatic potential and the 
reduced monomer concentration are plotted as function
of the position $x$ for a range of inter-surface distances $w$. 
As a result of screening the attraction of charged monomers 
to the surface is reduced considerably as compared to the 
low salt limit and the total amount of adsorbed polymer
is approximately half.
Despite the weaker adsorption, the qualitative behavior 
is similar.
 At short distances ($w\alt 15$\AA\ in Fig.~7b) 
a single adsorbed layer exists between the two surfaces.
At intermediate distances ($15$\AA$\alt w\alt 40$\AA) this layer 
separates into two strongly interacting layers
and at larger distances ($w\agt 40$\AA) the two layers decouple 
from each other and only weakly interact with each other.

When the separation is large enough ($w\agt 15$\AA\ in Fig.~7) 
the adsorption is strong enough so that 
the charged polymer over--compensates the surface charges.
The signature of this effect is that the electrostatic potential 
changes sign in the central region as seen in Fig.~7a.
This sign reversal strongly affects the small ion
concentration  $c^\pm(x)$ as demonstrated
in Fig.~8. 
Since the concentration of the small ions follows a Boltzmann 
distribution $c^\pm(\rr)=c^\pm_b \exp(\mp\beta e\psi)$
the concentration of small negative ions 
$c^-$ in the central region exceeds 
that of the positive ions $c^+$ 
as can be seen in Fig.~8 for $w=40$\AA\ (solid curve). 
Altogether, the net charge (per unit area) in the central region 
will be negative ensuring an overall charge neutrality
between the two surfaces.
Note, that at small separations (e.g. $w=10$\AA\ in Fig.~8) 
the potential is negative everywhere and the concentration of 
positive ions exceeds that of the negative ions everywhere 
within the gap.

%- - - - - - - - - - - - - - - - - - - - - - - - - - - - - -
\noindent{\bf Polyelectrolyte Adsorption:}~~~
  Despite the effect of screening, the adsorption of 
strongly charged polyelectrolytes is strong as can be seen in Fig.~7b. 
The reason is that screening has two competing effects: 
on one hand it reduces the attraction of charged monomers 
to the surface which is the driving force for adsorption.
On the other hand screening also reduces the monomer--monomer 
Coulomb repulsion between adsorbed monomers, thus allowing for
more charges to accumulate near the surface. 
Hence, despite the fact that the range of the electrostatic
interaction is reduced considerably, the average polymer 
concentration near the surface can be high (as long as $p$ is not too
small).

  In contrast with the low salt regime, here the small ions 
play an important role in balancing the surface charges. 
  In Fig.~9 the different contributions to the 
charge densities per unit area are plotted 
as function of the distance $w$.
The negative curve corresponds to the 
(induced) surface charge density on
the two surfaces
 $2\sigma_s$. 
The positive curves correspond to the total amount of 
polymer charges $\sigma_p$ (per unit area) 
which have adsorbed between the two surfaces,
and the 
total amount of charge carried by small ions $\sigma^++\sigma^-$.
At short distances, $w\alt6$\AA, 
the polymer contribution is small and the
main contribution to the charge density is that of the small ions.
This distance can also be regarded as a 
lower cutoff to the continuum theory employed here since the monomer
size we employed is of the same order of magnitude ($a=5$\AA).

   When the surfaces are taken further apart, the polymer contribution 
and the net contribution of the small ions saturate to a constant 
value. This occurs when the two adsorbing surfaces decouple 
from each other, and two distinct adsorbed layers build up on 
each of the surfaces. 
Unlike the low salt case where the contribution of the small ions 
is negligible and the charged polymers dominate the charge density, 
in the high salt regime the contributions of the small ions 
and the polymer are comparable in magnitude. 
 For example, in Fig.~9, the polymer contributes about one third 
of the charge density (per unit area) between the surfaces 
while the small ions contribute the other two thirds.

%- - - - - - - - - - - - - - - - - - - - - - - - - - - - - -
\noindent{\bf Free Energies and Forces:}~~~
Screening has a pronounced effect on the inter--surface interactions
as can be seen in Fig.~10, where 
the inter--surface interactions are plotted as function of the 
distance for a strong polyelectrolyte ($p=1$) 
at high salt concentration.
The general behavior here is similar to the low salt case as
is seen in Fig.~6.
As the amount of salt increases, 
the adsorption is reduced because of the electrolyte screening
and the attractive forces become substantially weaker. 
Similar effects were also observed experimentally in 
SFA experiments \cite{Dahlgren-etal,Dahlgren}.

Since the forces are related to the mid-plane values of $\psi$ 
and $\phi$ (eq.~\ref{force}), it is of interest to study 
 the mid-plane values of $\psi(0)$.
In the inset of Fig.~10 the mid--plane value
of the reduced electrostatic potential $y(0)=\beta e\psi(0)$
is plotted 
as a function of the inter--surface distance for the same 
profiles that are used in the calculation of the forces.
As can be seen also in Figs.~3 and 7, the mid-plane potential is 
negative at short distances, changes sign to become positive
and finally decays to zero.
The
inter--surface interaction changes from repulsion to attraction at 
about the same distance where the mid--plane electrostatic 
potential changes sign. 
   This observation can be explained
by examining the various contribution
to the local free energy at the mid--plane, eqs.~\ref{fpol}-\ref{fel}.
At the mid-plane the squared gradient terms in $\phi$ and
$\psi$ vanish. In addition, for strong polyelectrolytes the excluded
volume and chemical potential terms are very small. 
The force $\Pi(w)=-f(x=0)$ is dominated by two terms. 
\begin{equation}
  \beta\Pi \simeq - p \phi^2(0) y(0) 
                  + 2 c_b \left[\cosh(y(0)) - 1\right]
  \label{scalingforcey}
\end{equation}
The first term is the contribution of the 
charged monomers and changes sign
when $y(0)$ changes sign. The second is the (repulsive) osmotic
pressure of the small ions and
is proportional to $y^2(0)$ at small
values of $y(0)$. It is clear from the above that as long as the 
mid--plane potential is small enough, $y(0)\ll p\phi^2(0)/c_b$,
the pressure is governed by the first term and will
change sign from attraction to repulsion when $y(0)$ changes sign.
It now follows that for a positively charged polymer, 
negative (positive) mid--plane potentials lead 
to repulsion (attraction) in agreement with Fig.~10.
Since the potential changes sign when the adsorbed monomers
over--compensate the surface charge, 
we conclude that this over--compensation is responsible 
for the reversal in the sign of the interaction.

Weak polyelectrolytes ($p\ll 1$) do not adsorb as much as 
strong polyelectrolytes. 
The attractive forces are much weaker
and are easily overpowered by the double layer repulsion of the 
small ions.
Such an example is presented in Fig.~11, where the inter--surface
forces are calculated for a low charge fraction ($p=0.1$) 
at two high salt concentrations $c_b=0.25$M and $c_b=1$M.
In contrast with the case of strong polyelectrolytes, 
here the forces are repulsive over the whole distance range
and decay on a length scale of $\kappa_s^{-1}$.

%-----------------------------------------------------------
\section{Scaling Regimes}
%-----------------------------------------------------------

The fundamental difficulty in studying polyelectrolytes 
is due to the competition between short range interactions
such as the chain elasticity and excluded volume interactions, 
and the long range electrostatic interactions.
In a previous work \cite{macromols} we have studied 
polyelectrolyte adsorption to a single surface
by separating the two competing length scales:
(i) The adsorption length $D$, which characterizes the 
width of the adsorbed layer
and (ii) the electrostatic screening length 
$\kappa_s^{-1}= (8\pi l_B c_b)^{-1/2}$ assuming that 
$c_b\gg p\phi_b^2$. 
The screening length depends on the salt concentration,
while the adsorption length depends on both electrostatic and 
non-electrostatic properties.

  The two length scales can be separated in 
two limits: (i) the low salt regime $D\ll\kappa_s^{-1}$ 
and (ii) the high salt regime $D\gg\kappa_s^{-1}$.
The difference between the two regimes being the range 
of the electrostatic interactions.
  The main assumption in this approach is that the polymer 
profile near a single flat surface can be written in the form 
\begin{equation}
   \phi(x)=\sqrt{C}\  h\biggl( {x\over D} \biggr)
   \label{scaling_prof}
\end{equation}
where $h(z)$ is a dimensionless function 
normalized to unity at its maximum and
 $C$ sets the scale of polymer adsorption.
The free energy can then be expressed in terms of $D$ and $C$
while the exact form of $h(z)$ affects only the
numerical prefactors. 
Minimization of the free energy with respect to $D$ and $C$
gives the single surface adsorption length $D_1$ and the 
concentration scale $C_1$.

   When two surfaces interact with each other, the single 
surface profile is affected by the presence of the other surface. 
As a result the shape of the profile changes with
the separation $w$ as demonstrated in Figs.~3 and 7. For example,
at short distances the profile varies monotonically between the 
surface and the half plane, 
while at larger distances it becomes non-monotonous, 
until finally the two surfaces decouple from 
each other, and the adsorption to each surface 
reduces to the single surface behavior.

As in the single surface case, it is advantageous to separate 
the different length scales. 
First, we compare the single surface adsorption length $D_1$, 
with the inter--surface separation $w$.
At large separations $w/2\gg D_1$
the surfaces interact weakly and the polyelectrolytes 
recover the single surface profiles.
 On the other hand, at short inter--surface distances $w/2\ll D_1$,
the gap is too small for the polyelectrolytes to follow
the single surface profile.
In this limit the relevant length scale  
(eq.~\ref{scaling_prof}) is just $D=w/2$, since $w/2$ serves as
a lower cutoff for $D$.
When $w/2$ increases so that $w/2\simeq D_1$,
the profile becomes more complex and our main 
assumption is no longer valid. 
  Nevertheless, as demonstrated below this simplified picture
reproduces the main features that characterize the inter--surface 
interactions in those short distances, $w/2 \le D_1$.

Furthermore, 
the effect of screening can be taken into account by separating 
the screening length from the two other length scales.
Two opposite limits are considered 
(i) The low salt regime $\kappa_s^{-1}\gg D_1$ and
(ii) The high salt regime  $\kappa_s^{-1}\ll D_1$.

%- - - - - - - - - - - - - - - - - - - - - - - - - - - - - -
\subsection{Low Salt Regime: $\kappa_s^{-1}\gg D_1$}
%- - - - - - - - - - - - - - - - - - - - - - - - - - - - - -

  In the low salt regime the screening length is much 
larger than the width of the adsorbed layer and 
the effect of the small ions on the structure of the 
adsorbed layer can be neglected.
  This assumption amounts to neglecting the entropic contribution
to the free energy $f_{\rm ions}(\rr)$ (eq.~\ref{fent}) 
and the electrostatic energies of the small ions in 
 $f_{\rm el}(\rr)$ (eq.~\ref{fel}). 

%- - - - - - - - - - - - - - - - - - - - - - - - - - - - - -
\subsubsection{Large Distances: $w/2 \gg D_1$}
%- - - - - - - - - - - - - - - - - - - - - - - - - - - - - -

  At large distances $w/2\gg D_1$, the two surfaces are
only weakly coupled. 
The structure of the adsorbed layer near each of the two surfaces 
reduces to the single surface profile, and the ``decorated''
surfaces interact through a weak double layer interaction.
In the limit of large distances and low salt conditions one
needs to address the question of the relative size of $w/2$
and $\kappa_s^{-1}$ as both lengths are large.

The free energy of an isolated adsorbing surface
can be approximated by \cite{macromols}
\begin{equation}
  \beta F^{(1)}_p(C,D) = \alpha_1{a^2\over 6D}C - \alpha_2 p|y_s|C D
          + 4\pi \beta_1 l_Bp^2 C^2D^3
          + \frac{1}{2} \beta_2 v C^2 D
  \label{scalingF}
\end{equation}
  The first term is the polymer elastic energy (or connectivity) 
term, the second term is the electrostatic interaction of 
the monomers with the surface, and the third term is the Coulomb 
repulsion between the adsorbed monomers. 
The electrostatic terms can be derived by integrating the interaction 
of every pair of charged layers at distances $x$ and $x'$ 
from the surface, with charge densities (per unit area)
 ${\rm d}\sigma =pe\phi^2(x ){\rm d}x$ and 
 ${\rm d}\sigma'=pe\phi^2(x'){\rm d}x'$, respectively.
Finally, the last term is the excluded volume term and will be 
neglected here since at low salt concentration 
its contribution is important only 
for extremely weakly charged polyelectrolytes. 

The coefficients $\alpha_1,\alpha_2,\beta_1$ and $\beta_2$ are numerical
prefactors, which depend on the exact shape of 
the dimensionless scaling function $h(z)$.
These coefficients can be explicitly calculated for a specific profile 
by integrating the Poisson equation without taking into account
the small ion contributions. 
For the simplest monotonous profile, 
 namely a linear profile $h_1(z)=z$ for $0\le z\le 1$ 
 and $h_1(z)=0$ for $z> 1$, we get 
 $\alpha_1=1,\alpha_2=1/3,\beta_1=1/14$ and $\beta_2=1/5$.
For a non-monotonous parabolic profile,
 $h_2(z)=4z(1-z)$ for $0\le z\le 1$ and $h_2(z)=0$ elsewhere, we get 
 $\alpha_1=16/3,\alpha_2=8/15,\beta_1\simeq 1/9$ and $\beta_2\simeq 2/5$.
Another profile which we consider is an intermediate profile of 
the form $h_3(z;\eta)=4/\eta^2 z(\eta-z)$ for $0\le z\le 1$ 
and $h_3(z)=0$ elsewhere, where $\eta$ is a parameter. 
This profile is non-monotonous and has
 a finite value at $z=1$,
which corresponds to the symmetry plane between the two surfaces.
Furthermore, the special cases $\eta\gg 1$ and $\eta=1$
reduce to the simple linear and parabolic 
profiles $h_1(z)$ and $h_2(z)$, respectively.
The parabolic profile $h_2(z)$ is a good choice for 
an isolated adsorbing surface in contact with a bulk of low 
concentration, 
whereas $h_1(z)$ describes better interacting surfaces
at small separation. The third profile $h_3(z;\eta)$ 
can be regarded as intermediate between the other two.
We stress that our scaling results do not depend on
the specific shape of the profile $h(z)$. Only the numerical
prefactors will change.

 The single surface  free energy (eq.~\ref{scalingF}) can be
minimized with respect to both $D$ and $C$ along the same lines
as was done in ref.~\cite{macromols}.
This yields a length scale $D_1$ characterizing the adsorption 
onto a single surface
\begin{equation}
  D_1 \simeq {a \over p^{1/2}|y_s|^{1/2}} 
  \label{scalingD1}
\end{equation}
and a concentration scale 
\begin{equation}
  C_1 \simeq {|y_s|^2 \over l_B a^2}
  \label{scalingC1}
\end{equation}

In the low salt limit screening effects can be neglected 
as long as the screening length $\kappa_s^{-1}$
is much larger than the adsorption length $D_1$.
This condition limits the low salt regime to 
\begin{equation}
  \label{cbLS}
  c_b \ll {p |y_s| \over l_B a^2}
\end{equation}

Inserting the above expressions back in the free energy gives 
the single surface free energy (up to numerical factors):
\begin{equation}
  \beta F^{(1)}_p \simeq - {p^{1/2} |y_s|^{5/2} \over l_b a}
  \label{scalingF1}
\end{equation}

  At distances larger than the adsorbed layer 
$x>D_1$ the amount of 
polyelectrolytes is small and comparable to its (low) bulk value. 
Since $p\phi_b^2 \ll c_b$ (even in the low salt limit), 
the interaction at large distances can be simplified.
The system can be regarded as
a solution containing electrolytes  only 
(no polyelectrolytes) between two {\em effective} 
surfaces positioned at the 
edge of the adsorbed layers $x=\pm (w/2-D_1)$.
The effective inter--surface distance is now $w_{\rm eff}=w-2D_1$
and each surface is kept at a (reduced) potential 
$y_D=\beta e\psi_D$ which is much smaller in magnitude 
than the original surface potential $|\psi_D|\ll|\psi_s|$.

In the absence of polyelectrolytes in the effective gap,
the electrostatic potential 
between to charged surfaces can be obtained by solving the 
Poisson--Boltzmann equation \cite{IsraelachviliBook}:
\begin{equation}
   \nabla^2\psi(\rr)
   = \frac{8\pi e}{\varepsilon}c_b \sinh(\beta e\psi)
   \label{PB}
\end{equation}
The above equation can be obtained from eq.~\ref{PBs}
in the no polyelectrolyte limit.
After the differential equation has been solved with the 
appropriate boundary conditions (namely, $\psi=\psi_D$),
the repulsive free energies and inter--surface forces 
can be calculated.
In particular, eq.~\ref{PB} can be solved analytically 
in the weakly coupled regime, $\kappa_s^{-1}\ll w_{\rm eff}/2$,
yielding the following expressions
for the repulsive inter--surface interactions:
\begin{equation}
  \label{Fel}
   \beta\Delta F_{\rm el} 
    = 64 c_b \kappa_s^{-1} \tanh^2(y_D/4) \e^{-\kappa_s w_{\rm eff}}
    \simeq  16 c_b \kappa_s^{-1} y_D^2 \e^{-\kappa_s w_{\rm eff}} 
\end{equation}
\begin{equation}
     \label{PIel}
   \beta\Pi_{\rm el} 
    = 64 c_b \tanh^2(y_D/4) \e^{-\kappa_s w_{\rm eff}}
    \simeq 16 c_b y_D^2 \e^{-\kappa_s w_{\rm eff}}
\end{equation}

Other electrostatic regimes exist in which eq.~\ref{PB} can be solved
analytically. Those regimes lie beyond the scope of the present
study because the polymer adsorbed layer reduces substantially
the electric potential.
Unfortunately, our model is too simple to give an accurate estimate 
of $y_D$ which is a local property.
For this purpose more refined models are required.

%- - - - - - - - - - - - - - - - - - - - - - - - - - - - - -
\subsubsection{Short Distances; $w/2 \ll D_1 \ll \kappa_s^{-1}$}
%- - - - - - - - - - - - - - - - - - - - - - - - - - - - - -

 At short separations $w/2\ll D_1 \ll \kappa_s^{-1}$ 
the relevant length scale in eq.~\ref{scaling_prof} 
is $w/2$ instead of $D$. 
Consequently, and due to the planar symmetry of the system,
the polymer contribution 
to the free energy can be written as 
 $F_p(C,w) = 2 F^{(1)}_p(C,D=w/2)$ where 
 $F^{(1)}_p(C,D)$ is the single surface free energy 
(eq.~\ref{scalingF}).

The free energy is minimized now only with respect to $C$
leading to 
\begin{equation}
  \label{scalingC2}
   C \simeq {|y_s|\over l_B p}{w^2-w_{\rm min}^2 \over w^4}
\end{equation}
  where $w_{\rm min}^2= 2\alpha_1 a^2/3\alpha_2 p|y_s|$.
  The condition that $C$ be positive, limits the validity 
of eq.~\ref{scalingC2} to distances larger than
a minimal distance $w_{\rm min}$,
while at shorter distances $C=0$ and the polymers are depleted 
from the region between the surfaces. 
We estimate $w_{\rm min}\simeq 0.2 D_1$ and so eq.~\ref{scalingC2} 
is valid in the range $0.1 D_1 < w/2 < D_1$ \cite{wmin}.

The total amount of monomers (per unit area) adsorbed between the 
two surfaces is directly related to $C$ and $C_1$:
\begin{equation}
  \label{scalingGamma2}
   \Gamma(w) =  \int_{-w/2}^{w/2} \phi^2(x) \dx = 
     \left\{
     \begin{array}{ccrcl}
       0         &\hspace{1cm}&              &w& < w_{\rm min} \\
       \alpha_2 w C(w)  &\hspace{1cm}& w_{\rm min} <&w&< 2 D_1 \\
      2\alpha_2 D_1 C_1 &\hspace{1cm}&       2D_1  <&w&
      \end{array}
      \right.
\end{equation}

   Similarly, the average reduced monomer concentration is given by 
\begin{equation}
  \label{scaling_u2}
  \left< {\phi^2\over\phi_b^2} \right> = {\Gamma(w) \over w\phi_b^2} = 
     \left\{
     \begin{array}{ccrcl}
       0           &\hspace{1cm}&            &w& < w_{\rm min} \\
       \alpha_2{C(w)/\phi_b^2} &\hspace{1cm}&w_{\rm min} <&w&< 2 D_1 \\
      2\alpha_2 {D_1 C_1/ w\phi_b^2} &\hspace{1cm}&       2D_1  <&w&
      \end{array}
      \right.
\end{equation}

  The adsorption properties $\Gamma(w)$ and
 $\left<\phi^2/\phi_b^2\right>$
are plotted in Fig.~12 as function of the inter--surface 
distance $w$ for the same physical values of Fig.~4.
Our results agree qualitatively with the numerical results of 
Fig.~4 and reproduce the three different adsorption regimes.

  Inserting the above expression for $C$ (eq.~\ref{scalingC2})
 back into the free energy, yields:
\begin{equation}
  \label{scalingF2}
   \beta F_p \simeq - {|y_s|^2\over l_B}
                      {(w^2-w_{\rm min}^2)^2 \over w^5}
\end{equation}
for $w_{\rm min}<w< 2 D_1$. At distances shorter than $w_{\rm min}$, 
the polyelectrolyte is depleted from the gap, 
and the inter--surface interaction is dominated by electrostatic
repulsion \cite{depletion}. 

  The inter-surface force $\Pi_p$ is readily obtained by 
differentiating the free energy $F_p$ with respect to $w$
\begin{equation}
  \label{scalingPi2}
   \beta\Pi_p = -\frac{\delta(\beta F_p)}{\delta w} \simeq 
   - {|y_s|^2 \over l_B}
     {(w^2-w_{\rm min}^2)(w^2-5w_{\rm min}^2) \over w^6}
\end{equation}

A quantitative comparison with the numerical results
is shown in Fig.~13 where the physical parameters are the same 
as in Fig.~6.
 In Fig.~13 the polymer contribution to
the excess free energy 
 $\Delta F_p(w) = F_p(w)-2 F_p^{(1)}$ and to the
inter--surface force $\Pi_p$
are plotted as  functions of the inter--surface separation $w$.
The single surface free energy $F_p^{(1)}$ is  
calculated from eq.~\ref{scalingF1}.
Only small separations $w<2D_1$ are shown in the figure.
For $w>2D_1$ the shape of the
inter-surface profile is more complex and we do not have scaling arguments
relating the polymer profiles with the force. However, we expect the 
polymer contribution to be small. 
Comparing Figs. 13 with 6, we note that our scaling results 
(the characteristic length scale as well as  the 
characteristic force)  are in good  agreement 
with the numerical results for several $p$ values. 

%- - - - - - - - - - - - - - - - - - - - - - - - - - - - - -
\subsection{High Salt Regime: $\kappa_s^{-1}\ll D_1$}
%- - - - - - - - - - - - - - - - - - - - - - - - - - - - - -

 In the high salt regime the screening length is much 
smaller than the adsorption length $D_1$. The Coulomb
interactions between the charged monomers and the surface and 
between the monomers themselves decay exponentially with the
Debye-H\"uckel screening length
$\kappa_s^{-1}$. 

Our calculation is based on estimating the polymer contribution
to the forces as mediated by the small ions. One should bear in mind
that the contribution of the small ions to the forces
is no longer negligible and can explain the discrepancy
between the numerical (exact) and the scaling results. It is hard to get
an analytical estimate to the small ions contribution because their
concentration depends on the polymer profile via the electric potential.

%- - - - - - - - - - - - - - - - - - - - - - - - - - - - - -
\subsubsection{Large Distance: $w/2 \gg D_1$}
%- - - - - - - - - - - - - - - - - - - - - - - - - - - - - -

The free energy (eq.~\ref{scalingF}) can be generalized by 
introducing $\kappa_s^{-1}$ as a cut-off on the range of the 
electrostatic interactions (similar to what was done in 
Ref.~\cite{macromols} for the single surface case):
\begin{equation}
  \beta F_p(C,D) = \alpha_1{a^2\over 6D}C - \alpha_2 p|y_s|C\kappa_s^{-1} 
          + 4\pi \beta_1 l_Bp^2 \kappa_s^{-2}C^2 D
          + \frac{1}{2} \beta_2 v C^2 D
  \label{scalingFcb}
\end{equation}
The electrostatic cut-off appears in two places.
In the second term, only the first layers up to a distance
 $\kappa_s^{-1}$ from the surface interact with the surface
charges. In the third term, each layer 
interacts only with its neighboring layers in the range of
 $\kappa_s^{-1}$.
The numerical values of the prefactors 
 $\alpha_1,\alpha_2,\beta_1$ and $\beta_2$ 
can in principle be different from the low salt values. 
However, since the prefactors are only used in Fig.~14 to demonstrate 
the qualitative behavior we will arbitrarily set their values 
to be the same as in the low salt regime. 

Minimizing eq.~\ref{scalingFcb} with respect to $D$ and $C$ yields
\begin{equation}
  D_1 \simeq  {\kappa_s a^2\over p|y_s|}\sim {c_b^{1/2}\over p}
  \label{scalingD1cb}
\end{equation}
and
\begin{equation}
  C_1 \simeq
            { p^2 |y_s|^2/(\kappa_s a)^2 \over 
                    \beta_1 p^2/c_b + \beta_2 v}
  \label{scalingC1cb}
\end{equation}
which are now also functions of the salt concentration $c_b$
and $\kappa_s=(8\pi l_B c_b)^{1/2}$.
The condition that the screening length is much smaller than the 
adsorption length amounts to 
\begin{equation}
  \label{cbHS}
  c_b \gg {p |y_s| \over 8\pi l_B a^2}
\end{equation}
  in agreement with the boundary of the low salt regime
(eq.~\ref{cbHS}).

The single surface free energy is now:
\begin{equation}
  \beta F^{(1)}_p \simeq - {p^3 |y_s|^3 \kappa_s^{-3} 
                  \over {(\beta_1 p^2/c_b + \beta_2 v) a^2}}
  \label{scalingFcb1}
\end{equation}

  As in the low salt regime, the two adsorbed layers interact
electrostatically.
However, since  the screening length 
is much shorter than the adsorption length in the high salt regime, 
this interaction decays quite rapidly. We note that
the high salt regime can be further divided into two
sub-regimes depending on the ratio of the two terms in the
denominator of $C_1$ and $F_p^{(1)}$ as was discussed 
in Ref.~\cite{macromols}.

%- - - - - - - - - - - - - - - - - - - - - - - - - - - - - -
\subsubsection{Short Distances: $w/2 \ll D_1$}
%- - - - - - - - - - - - - - - - - - - - - - - - - - - - - -

  At short distances, the relevant 
length scale in the free energy eq.~\ref{scalingFcb} 
is $w/2$ instead of $D$ (as in the low salt regime). 
The free energy can be minimized 
with respect to $C$ yielding
\begin{equation}
  \label{scalingC2cb}
   C \simeq {p|y_s|\kappa_s^{-1}\over \beta_1 p^2/c_b + \beta_2 v}
         {w-w_{\rm min} \over w^2}
\end{equation}
where $w_{\rm min} = \alpha_1\kappa_sa^2/3\alpha_2p|y_s|$.
As in the low salt case, $C$ is positive only for 
 $w>w_{\rm min}\simeq 0.2D_1$.
At smaller separations the polyelectrolytes are depleted from the 
gap and the inter-surface interaction is dominated by the 
electrostatic repulsion.
We also note that the validity of eq.~\ref{scalingFcb} requires that
$w\gg 2\kappa_s^{-1}$.

The polymer free energy of interaction is now
\begin{equation}
  \label{scalingFcb2}
   \beta F_p \simeq - {p^2|y_s|^2\kappa_s^{-2} 
                       \over \beta_1 p^2/c_b + \beta_2 v}
                      {(w-w_{\rm min})^2 \over w^3}
\end{equation}
and the inter--surface force
\begin{equation}
  \label{scalingPicb2}
   \beta\Pi_p \simeq - {p^2|y_s|^2\kappa_s^{-2} 
                         \over \beta_1 p^2/c_b + \beta_2 v}
                      {(w-w_{\rm min})(w-3w_{\rm min}) \over w^4}
\end{equation}

  The qualitative 
behavior described by eqs.~\ref{scalingFcb2} and \ref{scalingPicb2}
is similar to that of the low salt regime
 (eqs.~\ref{scalingF2}, \ref{scalingPi2}).
The typical behavior for a strong polyelectrolyte 
is depicted in Fig.~14, where the same 
physical parameters of Fig.~10 are used. We note that the
quantitative 
agreement between the numerical (Fig.~10) and scaling (Fig.~14) results
is not as good as in the low salt limit. 
For example, the value of the minimum in the free energy is about three
times smaller in Fig.~14 as compared with Fig.~10. Also the variation
of $w$ at this minimum with the salt concentration is weaker
in the numerical results. 
As discussed above, the main source of discrepancy 
between the numerical and scaling results is the omission of the 
small ion contribution in the latter.

%-----------------------------------------------------------
\subsection{Discussion}
%-----------------------------------------------------------

   To summarize our results we present in Fig.~15, a schematic 
diagram of the different adsorption regimes.
The dashed lines mark the single surface adsorption regimes
in terms of the charge fraction $p$ and the salt concentration $c_b$.
Three adsorption regimes can be distinguished: 
\begin{itemize}
\item The low salt regime $c_b \ll p |y_s|/8\pi l_B a^2$.
\item The first high salt (HS I) regime 
   $c_b \gg p |y_s|/8\pi l_B a^2$
   with weak polyelectrolytes $p^2\ll vc_b$.
\item The second high salt (HS II) regime 
   $c_b \gg p |y_s|/8\pi l_B a^2$
   with strong polyelectrolytes $p^2\gg vc_b$.
\end{itemize}

The shaded area in Fig.~15 marks the region in 
parameter space where the polymer contribution to the
 inter--surface interaction is comparable to or larger than 
the pure electrostatic contribution.
The shaded area includes the low salt regime, 
a large portion of the HS II 
(high salt/strong polyelectrolyte) regime   
and a small portion of the HS I 
(high salt/weak polyelectrolyte) regime.
   The exact crossover lines depend, of course, on the numerical 
coefficients which are not included in our approximations.
Nevertheless, the qualitative picture can be deduced from 
the diagram.
  The different behaviors (as depicted previously)
can be demonstrated with the help of this diagram.
The filled circles in the low salt regime mark
the graphs of Fig.~6 and are well within the 
shaded area.
The filled squares on the right border of the diagram
 (at $p=1$) correspond to the graphs of Fig.~10
representing strong polyelectrolytes in the high salt 
regime.
At higher salt concentration the system is closer
to the boundary of the shaded area
and the polymer attraction is weaker.
Finally, when the ionic strength is high enough
the attractive contribution is too weak to be observed.
Weak polyelectrolytes in the high salt regime
belong to the top left side of the diagram 
outside of the shaded area.
In this regime, the electrolyte dominates the 
inter--surface interactions which are purely repulsive, 
as is indeed the case for the force curves of Fig.~11.
These curves are represented in Fig.~15 by filled triangles.

  In the following, we briefly summarize our findings 
in the different adsorption regimes. In Sec. VI the findings 
are compared  with experimental works 
which are reported in the literature.

%----------------
\noindent{\bf Low Salt Regime:}~~~
  In the low salt regime, the Debye--H\"uckel screening length 
is much larger than the width of the adsorbed layer. 
As a result the electrostatic interactions of the 
charged monomers with the surfaces and their interactions
with other monomers are unscreened. 
This leads to strong adsorption as can be seen from the numerical 
results shown in Figs.~3--4 and from the scaling results 
(eqs.~\ref{scalingGamma2}, \ref{scaling_u2}) shown in Fig.~12.
In addition, since the bulk concentration of the salt
and counter-ions, $c_b$ and $p\phi_b^2$ respectively, is small, 
the charge density in the solution between the two surfaces
is mainly due to the charged monomers. 
This is demonstrated in Fig.~5, where it is shown that the 
surface charges are balanced by the charged monomers.

At large distances the adsorbed polymer forms two distinct layers on the 
two surfaces. The amount of polymer adsorbed between the two surfaces 
saturates to a constant value (Fig.~4) which is approximately 
twice the single surface adsorbed amount.
As discussed in a preceding work \cite{macromols},
the width of the single surface adsorbed layer $D_1$ 
and the single surface adsorbed amount $\Gamma_1$ 
both scale as $p^{-1/2}$. 
The dependence on $p$ is due to the balance between 
the attraction of the monomers to the surface which is proportional to $p$
and the monomer--monomer Coulomb repulsion which is proportional to $p^2$.
The fact that the adsorbed amount decreases when the polymer charge 
increases reflects the energy barrier for bringing a large amount of 
charged monomers to the vicinity of the charged surface.
 
  The two layers start to overlap when the inter--surface 
distance is about twice the width of the single surface adsorbed layer 
 $w\simeq 2D_1$.
Below this distance the adsorbed amount slightly increases (Fig.~4)
and the two surfaces strongly attract each other (Fig.~6).
Our scaling approach recovers 
the increase in the adsorbed amount (eq.~\ref{scalingGamma2} and Fig.~12) 
and the attraction of the two surfaces 
(eqs.~\ref{scalingF2},\ref{scalingPi2} and Fig.~13).
The magnitude of the polymer contribution to the interaction free energy 
scales as
\begin{equation}
  \label{Fp_scale}
  \beta \Delta F_p \sim {p^{1/2} |y_s|^{1/2} \over l_B a}
\end{equation}
This energy scale should be compared with 
the electrostatic interaction energy which scales as
\begin{equation}
  \label{Fel_scale}
  \beta \Delta F_{el} \sim c_b \kappa_s^{-1} |y_s|^2 
\end{equation}
The condition that $\Delta F_p$ will be at least comparable to $\Delta F_{el}$
limits the salt concentration to
\begin{equation}
  \label{cbLSatt}
  c_b < {p |y_s| \over l_B a^2}
\end{equation}
in agreement with the boundary of the low salt regime (eq.~\ref{cbLS}).

If the inter--surface distance is further reduced, 
the entropy loss due to the confinement of the polymer 
to a narrow slit pushes the polymer out of the gap between 
the two surfaces. 
This can be seen from the numerical results (Fig.~4) and also 
from eq.~\ref{scalingC2}, where $w_{\rm min}$ is the minimal distance 
below which the polymer is compelled to leave the gap. 

%-------------------------
\noindent{\bf High Salt Regime:}~~~
In the high salt regime the Debye--H\"uckel screening length 
 $\kappa_s^{-1}$ is much smaller than the width of the adsorbed layer.
As a result the range of the electrostatic interactions is much shorter
and each charged monomers interacts only with monomers at a distance 
smaller than $\kappa_s^{-1}$.

The limiting behavior at large distances depends strongly on the 
charge fraction $p$.
For weak polyelectrolytes where $p$ is small (regime HS I) 
the monomer--monomer Coulomb repulsion which is proportional to $p^2$ 
is negligible and the single surface adsorbed amount $\Gamma_1$ 
scales as $p/c_b^{1/2}$.
On the other hand, for strong polyelectrolytes where $p$ is large 
(regime HS II) the monomer--monomer Coulomb repulsion is dominant 
and the single surface adsorbed amount $\Gamma_1$ 
scales as $c_b^{1/2}/p$.
The latter behavior is similar 
to that of the low salt regime with a 
different $p$ dependence. At higher salt concentration the 
adsorbed amount increases as the monomer--monomer Coulomb repulsion 
at the adsorbed layer is screened out. 

In the high salt case, the contribution of the 
small ions to the charge density can not be neglected. 
As seen in Fig.~9, about two thirds of the surface charge are balanced 
by small ions and only one third by charged monomers.
Another interesting aspect of the interplay between charged polymer
chains and small ions is the spatial distribution of charges between the 
two surfaces.
%
%Since each polymer chain carries a large number of charged monomers,
%its attraction to the surface is much stronger than that of a single ion.
%This causes the monomers to adsorb at a flat layer close to the surface.
%
The polyelectrolytes are strongly adsorbed on the surface
resulting in 
a sign reversal of the potential: it 
is negative at the surface and becomes positive at a distance of 
 $6-7$\AA.
In order that the system will be overall neutral the central region 
between the two surfaces has an excess of negative ions as seen in 
Fig.~8. 
At short distances (less than $12-14$\AA)
this effect disappears and the potential is negative
everywhere in the gap.
As seen from Fig.~10 and eq.~\ref{scalingforcey} the sign reversal of 
the mid--plane potential $y(x=0)$ is accompanied by a sign reversal 
in the inter--surface force $\Pi(w)$.
The potential sign reversal can also be seen in the low salt regime. 
Since the concentration of small ions is very small, 
the over-compensation effect is weak.

When the two surfaces are brought closer together,
$w<2D_1$, these layers start to overlap 
(See Fig.~7) and the adsorbed amount slightly increases.
At this separation, strong polyelectrolytes
induce strong attraction between the two surfaces (Fig.~10). 
The polymer contribution to the attraction can be estimated from 
our scaling approach to be 
\begin{equation}
  \label{Fpcb_scale}
  \beta \Delta F_p \sim {p^3 |y_s|^3 \kappa_s^{-3} \over 
                 ( \beta_1 p^2/c_b + \beta_2 v) a^2}
\end{equation}
Following the low-salt discussion we compare
this interaction with the electrostatic interaction
energy (eq.~\ref{Fel_scale}).
For weak polyelectrolytes $p^2 \ll vc_b$ (regime HS I) 
the polymer contribution dominates for
\begin{equation}
  \label{cbHSIatt}
  c_b^2 < {p^3 |y_s| \over l_B a^2 v}
\end{equation}
while for strong polyelectrolytes $p^2 \gg vc_b$ 
the polymer contribution
is dominant at low salt concentration
\begin{equation}
  \label{cbHSIIatt}
  c_b \ll {p |y_s| \over l_B a^2}
\end{equation}

At very short distances $w<w_{\rm min}$ (eq.~\ref{scalingC2cb}),
the polymer is depleted from within the gap as can be also seen in 
Fig.~9.

%-----------------------------------------------------------
\section{Irreversible Adsorption}
%-----------------------------------------------------------

  So far in this study we have assumed that the adsorbed 
layer is in thermodynamic equilibrium with a bulk reservoir.
Hence the total amount of adsorbed monomers can
vary and is determined by the free energy minimization 
(eqs.~\ref{F}-\ref{fel}).
However, in physical systems
the energetic barrier for detaching an adsorbed 
chain from the surface can be much larger than the 
thermal energy $k_BT$. 
As a result the relaxation times towards  
equilibrium can be much larger than the experimental 
time scales. In this case, one can consider the
amount of adsorbed monomers as fixed.

  In this section we study the effect of irreversible 
adsorption by excluding the possibility of polymer exchange 
with the reservoir while keeping the total amount of 
monomers adsorbed between the two surfaces fixed.
For simplicity we limit ourselves to the low--salt case. 
The results can be generalized to the high salt case.
  In principle, the state of the system is determined by 
minimizing the free energy (eqs.~\ref{F}-\ref{fel})
under the constraint that 
\begin{equation}
   \int_{-w/2}^{w/2} \phi^2(x) \dx = \Gamma_0
   \label{Gamma0}
\end{equation}
  where $\Gamma_0$ is the predetermined value of the 
amount of polymer between the surfaces.
This constraint can be introduced 
through a Lagrange multiplier $\lambda$ 
replacing the chemical potential term in $F$ so that the 
functional to be minimized becomes
\begin{equation}
   \beta \tilde{F} = \beta F 
       - \lambda \left(\int_{-w/2}^{w/2}\phi^2(x)\dx - \Gamma_0\right)
\end{equation}

The self--consistent
field equation now reads 
\begin{equation}
   \frac{a^2}{6} \nabla^2\phi(\rr)
   = v\phi^3 + p\phi\beta e\psi - \lambda\phi
   \label{SCFir}
\end{equation}
while the modified Poisson--Boltzmann equation (eq.~\ref{PBs}) is not 
affected by the irreversibility of the adsorption process, 
because the counter-ions are still free to exchange 
between the reservoir and the adsorbed layer.

   Equations~\ref{PBs} and \ref{SCFir} are solved under the 
constraint of eq.~\ref{Gamma0} where $\lambda$ is  adjusted
to give the desired value of $\Gamma_0$.
   Typical force profiles calculated numerically 
are presented in Fig.~16a,
where the free energy $2\pi\Delta F$ is 
plotted as a function of the inter--surface distance $w$.
For comparison we plot on the same graph the equilibrium
free energy (solid curve) of Fig.~6.
The dotted curve was calculated for the same physical values 
with the additional constraint that the
total amount of monomers adsorbed between the surfaces is fixed
to the equilibrium value of single surface adsorption  
(or equivalently two surfaces held at large distances).
This value is defined as $\Gamma_{\rm sat}$.
As can be seen from Fig.~4a for $p=1$,
$\Gamma_{\rm sat}=0.011$\AA$^{-2}$.

  The free energy difference  between 
the solid and dotted curves is quite small and appears 
only at short distances when the adsorbed amount 
in true equilibrium 
starts to deviate from its saturated value
(see also Fig.~4). Different values of $\Gamma_0$
are shown on Fig. 16a.
For low values of $\Gamma_0$ 
(e.g. $\Gamma_0=\Gamma_{\rm sat}/2$) the attraction is weaker,
 while for higher values 
(e.g. $\Gamma_0=2\Gamma_{\rm sat}$) the attraction is much 
stronger.

In order to understand better this behavior we return to the 
scaling approximation where we consider first the case of 
irreversible adsorption on a {\it single} surface. 
We assume that a fixed amount $\Gamma=\Gamma_0/2$
is adsorbed to the surface.
Since the adsorbed amount $\Gamma$ is related to the 
length scale $D$ and the concentration scale $C$ through 
 $\Gamma=\alpha_2 CD$, it is possible to express 
$C$ in terms of $D$ and thus the 
free energy (eq.~\ref{scalingF}) in terms of $D$ and $\Gamma_0$.
Neglecting the excluded volume term, equation~\ref{scalingF} 
now becomes:
\begin{equation}
  \beta F^{(1)}_p(\Gamma_0,D) 
    = {\alpha_1\over 12\alpha_2}{a^2\over D^2}\Gamma_0 
    - {1\over2}p|y_s|\Gamma_0
    + {\pi\beta_1\over\alpha_2^2} l_Bp^2\Gamma_0^2 D
  \label{scalingFir}
\end{equation}

Minimization of the free energy with respect to $D$ 
yields
\begin{equation}
  D_{\rm irr} \simeq \left({a^2\over l_B p^2\Gamma_0}\right)^{1/3}
  \label{scalingDir1}
\end{equation}
Substituting $D_{\rm irr}$ back in the free energy gives
\begin{equation}
  \beta F^{(1)}_p \simeq - {1\over2}p|y_s|\Gamma_0       
    + a^{2/3} l_B^{2/3} p^{4/3} \Gamma_0^{5/3}
  \label{scalingFir1}
\end{equation}
  where we have omitted some numerical coefficients.
  The first term in eq.~\ref{scalingFir1} is simply 
the interaction energy of the charged monomers with the surface 
while the second term is a balance between the monomer--monomer
Coulomb repulsion and the chain elasticity term.

When the two surfaces are interacting with each other,
we can write an explicit expression for $F_p(w)$ 
at small distances ($w\ll D_{\rm ir}$) by replacing the 
adsorption length scale $D$ with $w/2$.
The free energy then becomes:
\begin{equation}
  \beta F_p(w) 
    = {2\alpha_1\over 3\alpha_2}{a^2\over w^2}\Gamma_0 
    - p|y_s|\Gamma_0       
    + {\pi\beta_1\over\alpha_2^2} l_Bp^2\Gamma_0^2 w
  \label{scalingFir2}
\end{equation}

   The interaction free energy is now $\Delta F_p=F_p(w)-2F^{(1)}_p$.
The surface interaction term cancels out and we are left with 
three terms: two positive terms from eq.~\ref{scalingFir2}
and a negative term from eq.~\ref{scalingFir1}.
The latter scales as $\Gamma_0^{5/3}$ and is responsible to the 
attraction at short distances where the third term of 
eq.~\ref{scalingFir2} becomes small.
In Fig.~16b we plot the interaction free energy as calculated from 
eqs.~\ref{scalingFir1},\ref{scalingFir2} for the same 
physical values as in Fig.~16a. 
Although the numerical coefficients can not be obtained 
accurately from this approach the qualitative behavior 
is in accord with the numerical results.

%-----------------------------------------------------------
\section{Comparison With Experiments}
%-----------------------------------------------------------

   The experimental studies with the Surface Force Apparatus 
\cite{Klein,Marra,Claesson,Dahlgren-etal,Audebert,Eriksson,Dahlgren,Helm}
focus mostly on the repulsive interactions between adsorbing 
surfaces. Due to the limitations of the experimental technique, 
the attractive interactions 
at short distances appear as jumps in force--distance 
profiles. 
Nevertheless, some of the qualitative features can be 
deduced from the experiments and agree with our findings.

  {\bf Luckham and Klein} \cite{Klein} have measured interactions
between mica surfaces in the presence of poly-L-lysine which 
is a strong polyelectrolyte ($p=1$) 
at two different salt concentrations $c_b=1$mM and $c_b=0.1$M.
The inter--surface forces $\Pi_R(w)$ (eq.~\ref{SFA}) were measured
at distances $50$\AA\ $<w<1200$\AA.
The forces were always repulsive, decaying exponentially as 
function of $w$ with decay lengths comparable 
to the Debye--H\"uckel screening length $\kappa_s^{-1}$.
These forces can be interpreted as the electrostatic 
repulsion between two adsorbed layers at distances 
larger than the width of a single layer (eq.~\ref{Fel}).
Significant deviations were found between the first approach 
where the two surfaces are brought close together and
subsequent decompression--compressions cycles.
The amplitude of the repulsion in the latter case was strongly reduced 
while the decay length of the force remained quite the same. 
In addition, the amount of polymer adsorbed between the two surfaces,
as estimated by refractive index measurements, was much higher
than in the initial measurements.
Those effects demonstrate that the adsorbed layers are not always in 
equilibrium and that compression might lead to strong adsorption
of charged polymers. 
The adsorbed amount remains high when the surfaces are 
separated from each other due to the high energetic barrier
for desorption.
The reduction in the inter--surface repulsion when the adsorbed 
amount is high is in accord with our numerical and analytical 
results for the case of irreversible adsorption (Sec. V).

  {\bf Marra and Hair} \cite{Marra} have adsorbed 
poly(2-vinylpyradine) (P2VP) between mica surfaces. 
The pH of the solution was such that the polymer was fully 
charged ($p=1$) during the experiments.
At low salt concentrations the forces were repulsive 
at large distances with an exponential decay which is 
consistent with the Debye--H\"uckel screening length.
Attraction of about $-7$mN/m was detected at distances
between $11$ and $40$\AA.
In the presence of $0.01$M NaCl the magnitude of 
the repulsive forces increased 
while the attraction was reduced to $-3$mN/m
and shifted to distances between $25$ and $80$\AA.
At even higher ionic strengths ($0.1$M NaCl) 
the attraction disappeared altogether
but an additional non--exponential contribution to the 
inter--surface repulsion was detected at distances between 
$60$ to $100$\AA.
These results confirm our findings that the effect of salt 
is to increase the adsorption length $D_1$ and to reduce 
the polymer attractive contribution to the inter--surface interactions.

{\bf Claesson and Ninham} \cite{Claesson} have adsorbed chitosan,  
a cationic biopolymer of glucosamine segments, 
between mica surfaces 
in the presence of $0.01$ wt \% acetic acid. 
The chitosan charge fraction was controlled through the pH of 
the solution
\begin{equation}
  p = {10^{-({\rm pH-pK}_0)} \over 1 + 10^{-({\rm pH-pK}_0)} }
  \label{pHpK}
\end{equation}
where ${\rm K}_0 = 10^{-\rm p K_0}\simeq 10^{-6.5}$ is the dissociation
constant
of the chitosan monomers.
At low pH, the polymer is fully charged and the interactions 
were repulsive in the first compression when
the surfaces were brought into contact.
A double layer repulsion was detected at large distances ($w>100$\AA)
and strong steric repulsion at shorter distances.
Upon separation, attraction was detected at distances around 
 $20-25$\AA.
At pH=$6.2$  ($p\simeq 2/3$) the repulsive double layer
interaction disappeared altogether. 
The disappearance of the electrostatic double layer interaction
indicates that the surface charges are exactly balanced by the 
adsorbed polymers.
Attraction was detected at distances of about $20-25$\AA\, 
and strong steric repulsion at shorter separations.
At pH=$9.1$ where the polymer was only weakly charged ($p\simeq 1/400$)
the double layer repulsion was again the dominant interaction.

  {\bf Dahlgren et al.} \cite{Dahlgren-etal} have studied the 
effect of salt concentration by adsorbing 
poly ((3-methacrylamido propyl) -trimethyl-ammonium chloride) 
(MATPAC) between mica surfaces.
Three salt concentrations were considered $c_b=0.1$mM, 
 $0.01$M and $0.1$M.
The forces were repulsive at large distances and decayed exponentially
with decay lengths of $170$\AA, $30$\AA\ and $11$\AA, respectively.
The first decay length is smaller than the Debye--H\"uckel screening
length of the corresponding salt concentration 
($\kappa_s^{-1}\simeq 300$\AA) due to the contribution of the 
counter-ions. At higher salt concentrations 
this contribution is negligible and the decay lengths agree 
with the expected screening lengths.
Attractive interactions were detected at short distances of a 
few nanometers.
The magnitude of these attractive forces decreased as the amount of 
salt increased in agreement with the numerical results of Fig.~10.

In another work, {\bf Dahlgren et al.} \cite{Audebert} 
have studied the effect 
of both charge fraction and salt concentration
on the inter--surface forces.
Three different polyelectrolytes with different charge fractions $p$ 
were used: MAPTAC ($p=1$) and 
two copolymers AM-CMA-10 ($p=0.1$) and AM-CMA-30 ($p=0.3$) 
which were prepared using different ratios of  
neutral acrylamide (AM) segments and positively charged 
(2-acroyloxy ethyl) -trimethyl-ammonium chloride (CMA) segments.
For a fixed charge fraction ($p=0.3$) three different 
ionic strengths were compared: $c_b=0.1$mM, $c_b=0.01$M 
and $c_b=0.1$M.
These experiments correspond to a vertical scan in Fig.~15.
At the lowest ionic strength, the system is in the low salt regime
and strong attraction is detected at intermediate 
distances $40<w<100$\AA.
At the next ionic strength ($c_b=0.01$M) the system is in the 
lower part of the high salt regime and weak 
attraction is still observed at distances below $60$\AA.
At the higher value of $c_b=0.1$M no attractive interactions 
are observed and the electrostatic repulsions dominate the 
inter--surface forces.
For a fixed ionic strength ($c_b=0.1$mM) Dahlgren et al. have 
compared the inter--surface interactions for three different 
charge fractions: $p=0.1$, $p=0.3$ and $p=1$. 
This set of experiments corresponds to a horizontal cut in Fig.~15.
At the lowest charge fraction the repulsive interactions
are dominant while for higher values of $p$ attractions is observed 
at distances below $w\simeq 100$\AA.

Finally, the effect of ionic strength was 
studied separately by {\bf Dahlgren} \cite{Dahlgren}.
 Two polyelectrolytes, 
poly (2-proplyionyloxy ethyl) -trimethyl-ammonium chloride) (PCMA)
 and MAPTAC which have different molecular weight but 
the same charge fraction ($p=1$) were studied.
The monomers of these polymers are large and therefore the regime 
boundaries in Fig.~15 should be shifted to lower salt concentrations.
Dahlgren has compared several types of multivalent salts 
at intermediate ionic strengths which are equivalent to 
 $c_b=0.1, 0.2, 0.3$ and $0.6$mM. 
In the higher ionic strengths ($0.3$ and $0.6$mM) no attractive 
interactions were observed. However, at lower ionic strengths 
attraction was observed at distances below $w\simeq 120$\AA\ and 
 $w\simeq 180$\AA\ for $c_b=0.1$mM and $c_b=0.2$mM, respectively.
The ratio between the two lengths scales is approximately 
 $\sqrt{2}\simeq 1.4$ in agreement with the adsorption length scale 
in the high salt regime $D_1\sim c_b^{1/2}$ (eq.~\ref{scalingD1cb}).
Furthermore, the adhesion force was measured as 
 $\Pi_{\rm ad}\simeq 170$ mN/m and $\Pi_{\rm ad}\simeq 250$ mN/m
in agreement with the force scale of strong polyelectrolytes
in regime HS II $F_p\sim 1/c_b^{1/2}$ (eq.\ref{Fpcb_scale}).

%-----------------------------------------------------------
\section{Conclusions}
%-----------------------------------------------------------

   In this paper we calculated numerically
concentration profiles of polyelectrolytes
between two charged surfaces
and studied the inter--surface 
interactions as a function of the distance between the charged 
surfaces. 
Over--compensation of surface charges by adsorbed monomers 
was found to be strongly related to the 
reversal of inter--surface forces from repulsion to attraction
at short distances where the two adsorbing layers strongly 
overlap.

  The effect of the polymer charge and ionic strength 
on the inter--surface interaction is studied by means of a
simple variational approach. 
  Three main regimes are found: 
(i) a low salt regime, $c_b \ll p |y_s|/8\pi l_B a^2$;
(ii) a high salt regime  $c_b \gg p |y_s|/8\pi l_B a^2$
 for weak polyelectrolytes $p^2\ll vc_b$ (HS I);
 and (iii) a second high salt regime 
 for strong polyelectrolytes $p^2\gg vc_b$ (HS II).

In the low salt regime, strong repulsion at very short distances 
is a result of the polymer depletion from the inter--surface gap.
As the distance increases to  $w\sim a/p^{1/2}$,  
strong attraction is due to overlap of the adsorbed layers.
Finally, when the inter--surface separation 
is larger than twice the adsorption length of a single surface, 
the two adsorbed layers separate and repel each other 
electrostatically.
In the HS II regime the behavior is similar to the low salt one,
with a modified length scale of interaction given by 
 $\kappa_s a^2/p$. On the other hand,
in the HS I regime, 
the polyelectrolyte attractive contribution 
is too weak to generate a similar
attraction at short distances. 
Consequently, the inter--surface interaction 
is repulsive with a decay length of $\kappa_s^{-1}$.

  Some of the features described above are also present 
in the discrete lattice model of B\"{o}hmer et al. \cite{Bohmer}.
In particular, attractive interactions between equally 
charged surfaces were obtained numerically 
(Fig.~9 of Ref.~\cite{Bohmer}).
This attraction was attributed to bridging by polymer chains 
from one surface to the other. 
The lattice model contains information regarding the fine 
details of the polymer chains which are absent in our model.
On the other hand, the continuum approach is a convenient 
starting point for analytical approximations such as the 
scaling approach presented in this work.
Attractive interactions were also obtained by 
Podgornik \cite{Podgornik} for the case of fixed adsorbed amount
and without considering the nonlinear excluded volume interaction.
For polyelectrolytes in a poor solvent, 
Ch\^atellier and Joanny \cite{Chatellier} have obtained 
oscillations in the polymer concentration as well as in the 
inter--surface interactions.

The model presented here takes into account the important
Coulombic degrees of freedom 
within the frame work of the Poisson-Boltzmann
formalism. We solve the coupled non-linear equations for 
the electrostatic potential and polymer concentration which
allows consistent treatment of excluded volume effects
as well as strong potential and surface charges (not
the linearized Debye-H\"uckel version). This allows us to
consider cases where the Coulombic degrees of freedom are a major
perturbation on the adsorption of neutral polymers.

In the same time our approach
has several limitations, some of which can be improved.
The polymers chains are treated
within a mean-field approximation which misses certain
properties of polymer
statistics such as the 
chain connectivity, 
stiffening of the persistence length, finite chain corrections
and more specific conformations of polymers close to surfaces (loops,
tails and trains). 

On the other hand, the simple model
we present offers a qualitative picture of polyelectrolyte chains
between surfaces and suggests several scaling regimes.
It can be further improved to take into account more realistic
situations such as surface heterogeneities and other geometries,
various charge distributions (quenched and annealed) on the chains
\cite{epl,epr,Raphael},
pH effects for acidic and basic polyelectrolytes \cite{macromols}
as well as finite ion
or monomer sizes \cite{Eigen,Iglic,prl}.

  In this work we have assumed constant surface potentials.
One could also consider constant surface charges. In the low
salt limit the behavior is expected to resemble the case of
fixed amount of adsorbed polymer since the charged monomers 
are main source of charges which are able to neutralize the 
surface charges. In the presence of salt this is no longer the 
case and the behavior can differ considerable.
 
It would be interesting to have thorough 
experimental results 
on the effect of the 
charge fraction $p$ and salt concentration $c_b$ on 
the nature and magnitude of the forces and the 
corresponding length scales. 
We hope that our present work 
will encourage such systematic experimental 
studies.

%-----------------------------------------------------------
% Acknowledgments
%-----------------------------------------------------------
\vspace*{0.5cm}
{\em Acknowledgments}

   We would like to thank P. Claesson, H. Diamant, 
B. J\"onsson, Y. Kantor, D. Langevin 
and S. Safran for useful discussions.  
   Two of us (IB and DA) would like to thank 
the Service de Physique Th\'eorique (CE-Saclay) 
and one of us (HO) 
the School of Physics and Astronomy (Tel Aviv University)
for their warm hospitality.
   Partial support from the Israel Science Foundation 
founded by the Israel Academy of Sciences and Humanities 
-- centers of Excellence Program 
and the U.S.-Israel Binational Foundation (B.S.F.) 
under grant No. 94-00291 gratefully acknowledged. 

%-----------------------------------------------------------
% References
%-----------------------------------------------------------
%\bigskip
%\section*{References}

%-----------------------------------------------------------
\pagebreak
\bigskip
\section*{Figure Captions}
\pagestyle{empty}
%-----------------------------------------------------------

\bigskip \noindent {\bf Fig.~1}:
Schematic view of a polyelectrolyte solution 
between two parallel charged surfaces
at a distance $w$ from each other.
The solution contains polyelectrolyte chains and 
small ions. The surfaces are kept at a constant potential.

\bigskip \noindent {\bf Fig.~2}:
(a)  Two half cylinders with a $90^\circ$ tilt
between the axes, as used in the surface force apparatus
to measure inter--surface forces. 
The radii $R$ of the cylinders are of the order 
of $1-2$ cm while $w$ ranges down to a few Angstroms.
(b) Two spheres of radii $R$ at a distance $w$. Typically, 
colloidal suspensions contain particles whose radii are of 
a few microns down to hundreds of Angstroms, 
while the stability is determined by the 
balance of forces at much smaller distances.

\bigskip \noindent {\bf Fig.~3}: 
 Profiles of 
(a) the reduced electrostatic potential $y=\beta e\psi$
and (b) the reduced monomer concentration $\phi^2/\phi_b^2$
as functions of the position $x$ between the two surfaces.
The profiles were obtained by solving numerically the differential 
equations (eqs.~\ref{PBs},\ref{SCFs}) 
for several inter--surface distances.
For comparison, the different profiles are plotted on the 
same axis so that all mid-planes ($x=0$) coincide. 
 The surfaces are placed at different distances from the mid-plane
as indicated by the filled squares.
In the numerical examples in Figs.~3 to 11
we assume the following physical parameters:
the polymer concentration is $\phi_b^2=10^{-6}$\AA$^{-3}$
with an effective monomer length $a=5$\AA\ 
and excluded volume parameter $v=50$\AA$^3$.
It is immersed in an aqueous solution ($\varepsilon=80$) 
at room temperature ($T=300$K) and 
the surfaces are kept at a constant potential 
 $y_s = \beta e\psi_s =-2$.
In addition, in the 
current figure the polymer charge fraction $p=1$ and 
the salt concentration $c_b=1$mM.
The different curves correspond to separations of
 $w=40$\AA\ (solid curve);
 $w=20$\AA\ (dots);
 $w=15$\AA\ (short dashes)
and  $w=10$\AA\ (long dashes).

\bigskip \noindent {\bf Fig.~4}:
 Adsorption of polyelectrolytes in a low salt solution 
between two charged surfaces for two different polymer charge
fractions. 
(a) Total amount of monomers adsorbed between the surfaces 
per unit area $\Gamma$ 
and (b) the average reduced monomer concentration 
$\langle\phi^2/\phi_b^2\rangle$
as  functions of the inter--surface distance $w$. 
  The salt concentration is $c_b=10^{-6}$M.
  The different curves correspond to charge fractions of
 $p=1$ (solid curve) and  
 $p=0.2$ (dashed curve).

\bigskip \noindent {\bf Fig.~5}:
   Charge densities per unit area 
as function of the inter--surface distance $w$. 
The two positive (upper) curves correspond to the total amount of 
polymer charge adsorbed between the two surfaces per unit area
 $\sigma_p= pe\Gamma $.
The two negative (lower) curves correspond to the induced surface 
charge density  on both surfaces, $2\sigma_s$.
The contribution of the small ions to the charge density is not 
displayed.
The physical parameters and notations are the same as in Fig.~4.

\bigskip \noindent {\bf Fig.~6}:
  Inter--surface interactions for polyelectrolytes
between two surfaces held at a constant potential and 
in a low salt solution. 
(a) The excess free energy per unit area $2\pi\Delta F$ 
as a function of the inter--surface distance $w$. 
The factor $2\pi$ is used in order to enable
direct comparison with SFA measurements
 (see eq.~\ref{SFA}).
(b) The force per unit area $\Pi$ between the two surfaces
as a function of the inter--surface distance $w$. 
The physical parameters and notations are the same as in Fig.~4.

\bigskip \noindent {\bf Fig.~7}: 
 Profiles of 
(a) the reduced electrostatic potential $y=\beta e\psi$
and (b) the reduced monomer concentration $\phi^2/\phi_b^2$
as function of the position $x$ between the two surfaces.
Same physical values and notations as in Fig.~3 except for a
much higher value for the salt concentration, $c_b=1$M.
The different curves correspond to separations of
 $w=40$\AA\ (solid curve);
 $w=20$\AA\ (dots);
 $w=15$\AA\ (short dashes);
 $w=10$\AA\ (long dashes)
and $w=7$\AA\ (dots and long dashes).

\bigskip \noindent {\bf Fig.~8}: 
  Profiles of the small ion concentration $c^+$ and $c^-$ 
as function of the position $x$ between the two surfaces 
for two of the separations presented in Fig.~7:
 $w=40$\AA\ (solid curve) and $=10$\AA\ (long dashes).

\bigskip \noindent {\bf Fig.~9}:
   Different contributions to the total  
charge density per unit area
as function of the inter--surface distance $w$ 
for highly charged polyelectrolytes ($p=1$)
in the high salt limit ($c_b=1$M).
The different curves correspond to
twice the induced surface charge density $2\sigma_s$, 
the total amount of polymer charge adsorbed between the 
two surfaces per unit area $\sigma_p$, 
and the total amount of charge carried by small ions 
($\sigma^++\sigma^-$).

\bigskip \noindent {\bf Fig.~10}: 
   Inter--surface interactions for highly charged polyelectrolytes
($p=1$) at high salt concentration. 
(a) The excess free energy per unit area $2\pi\Delta F$
and (b) the force per unit area $\Pi$ between the two surfaces
as function of the inter--surface distance $w$. 
The salt concentration is $c_b=0.1$M (solid curve)
and $c_b=1$M (dashed curve).
The inset shows the mid--plane values of the 
reduced electrostatic potential $y(0)$
as a function of $w$.

\bigskip \noindent {\bf Fig.~11}: 
   Inter--surface interactions for weakly charged polyelectrolytes
($p=0.1$) at high salt concentration. 
(a) The excess free energy per unit area $2\pi\Delta F$
and (b) the force per unit area $\Pi$ between the two surfaces
as function of the inter--surface distance $w$. 
The salt concentration is $c_b=1$M (solid curve)
and $c_b=0.25$M (dots).

\bigskip \noindent {\bf Fig.~12}: 
 Adsorption of polyelectrolytes in the low salt regime
as calculated from eqs.~\ref{scalingGamma2},\ref{scaling_u2}.
(a) Total amount of monomers adsorbed between the surfaces 
per unit area $\Gamma$, and
(b) the average reduced monomer concentration 
$\langle\phi^2/\phi_b^2\rangle$
as function of the inter--surface distance $w$.
The physical values and notations are the same as in Fig.~4.
The numerical prefactors of the intermediate profile $h_3(z,\eta)$
with $\eta=3/2$ were used in the calculation.
The vertical lines denote the distance where $w=2D_1$.

\bigskip \noindent {\bf Fig.~13}: 
   Polyelectrolyte contribution to 
(a) the interaction free energy, $2\pi\Delta F_p$ 
(eq.~\ref{scalingF2}), 
and (b) the inter-surface force, $\Pi_p$ (eq.~\ref{scalingPi2}),
in the low salt regime 
as function of the inter-surface separation $w$.
  Same physical values, notations and units as in Fig.~6. 
 The numerical prefactors of the intermediate profile $h_3(z,\eta)$
with $\eta=3/2$ were used in the calculation of the interaction
free energy and forces,
and the numerical prefactors of the parabolic profile
were used in the calculation of the single surface free energies.

\bigskip \noindent {\bf Fig.~14}:
   Polyelectrolyte contribution to 
(a) the interaction free energy, $2\pi\Delta F_p$ 
(eq.~\ref{scalingFcb2}), 
and (b) the inter-surface force, $\Pi_p$ (eq.~\ref{scalingPicb2}),
in the high salt regime 
as function of the inter-surface separation $w$.
  Same physical values, notations and units as in Fig.~10. 
 The numerical prefactors of the intermediate profile $h_3(z,\eta)$
with $\eta=3/2$ were used in the calculation of the interaction
free energy and forces,
and the numerical prefactors of the parabolic profile
were used in the calculation of the single surface free energies.

\bigskip \noindent {\bf Fig.~15}:
  Schematic diagram of the different regimes
as function of the charge fraction $p$
and the salt concentration $c_b$.
Three regimes can be distinguished:
(i) the low salt regime $D_1\ll\kappa_s^{-1}$;
(ii) the high salt regime (HS I) $D_1\gg\kappa_s^{-1}$
for weak polyelectrolytes $p\ll ({c_b v})^{1/2}$
and (iii) the high salt regime (HS II) $D_1\gg\kappa_s^{-1}$
for strong polyelectrolytes $p\gg ({c_b v})^{1/2}$.
The shaded area denotes the region where the polymer interactions 
are strong enough so that inter--surface attraction can be observed.
The filled circles correspond to the numerical force--distance 
profiles of Fig.~6, filled squares correspond 
to the numerical profiles of Fig.~10
and the filled triangles to the profiles of Fig.~11.

\bigskip \noindent {\bf Fig.~16}:
  The effect of irreversible polyelectrolyte adsorption 
on the interaction free energy.
The excess free energy per unit area $2\pi\Delta F$ is plotted 
as a function of the inter--surface distance $w$.
The graphs in (a) were obtained by solving numerically the 
mean field equations~\ref{PBs},\ref{SCFir} under the constraint 
of eq.~\ref{Gamma0}.
The solid curve corresponds to the equilibrium interaction 
free energies and is the same as the solid curve of Fig.~6.
The four curves were obtained for the same physical values
as in Fig.~5 with $p=1$ while the total amount of adsorbed 
monomers was kept at a constant value $\Gamma_0$.
The different curves correspond to 
 $\Gamma_0=\Gamma_{\rm sat}=0.011$\AA$^{-2}$ (dots),
 $\Gamma_0=2\Gamma_{\rm sat}$ (small dashes),
 $\Gamma_0=3\Gamma_{\rm sat}$ (long dashes) and
 $\Gamma_0=\Gamma_{\rm sat}/2$ (dot--dash line).
The graphs in (b) were calculated from the analytical 
expressions of the scaling approach 
(eqs.~\ref{scalingFir1},\ref{scalingFir2}).

\pagebreak \vfill

\begin{figure}[tbh]
  {\Large Fig.~1}
  \bigskip\bigskip\bigskip

  \epsfxsize=0.5\linewidth
  \centerline{\hbox{ \epsffile{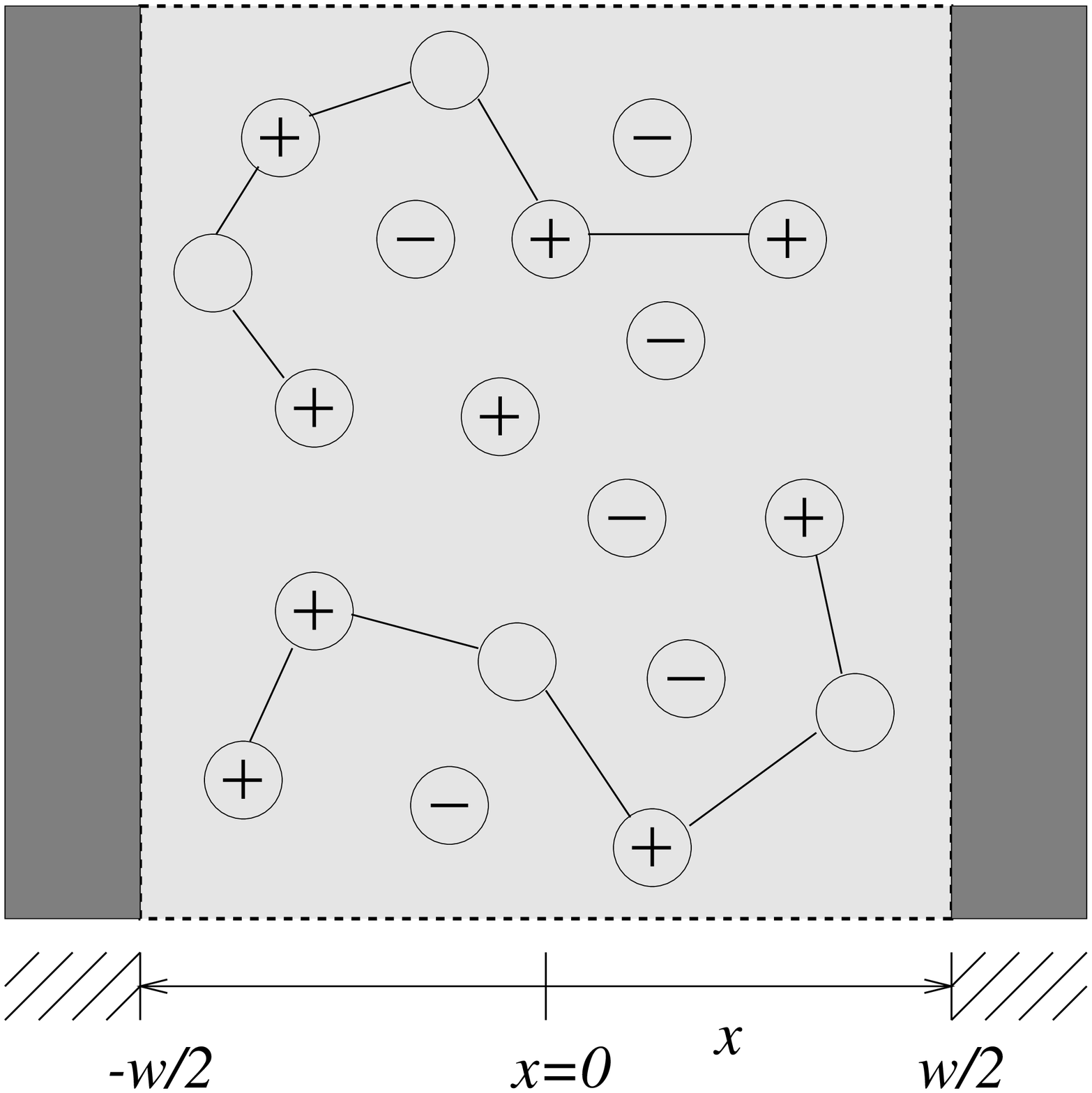} }}
\end{figure}

\begin{figure}[tbh]
  {\Large Fig.~2}
  \bigskip\bigskip\bigskip

  \epsfysize=16\baselineskip
  \centerline{\hbox{ \epsffile{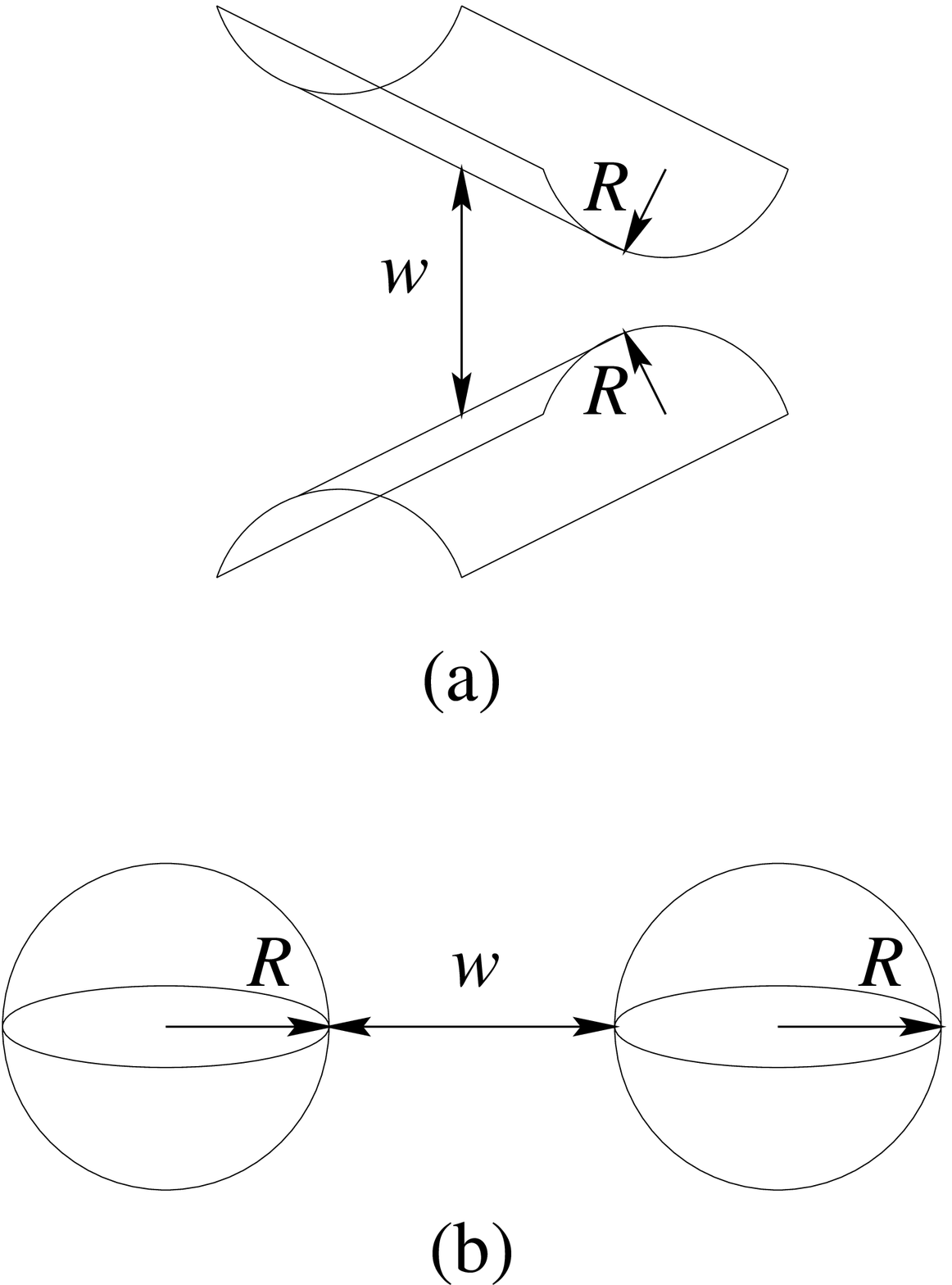} }}
\end{figure}
\vfill

\pagebreak \vfill

\begin{figure}[tbh]
  {\Large Fig.~3}
  \bigskip\bigskip\bigskip

  \epsfxsize=0.5\linewidth
  \centerline{\hbox{ \epsffile{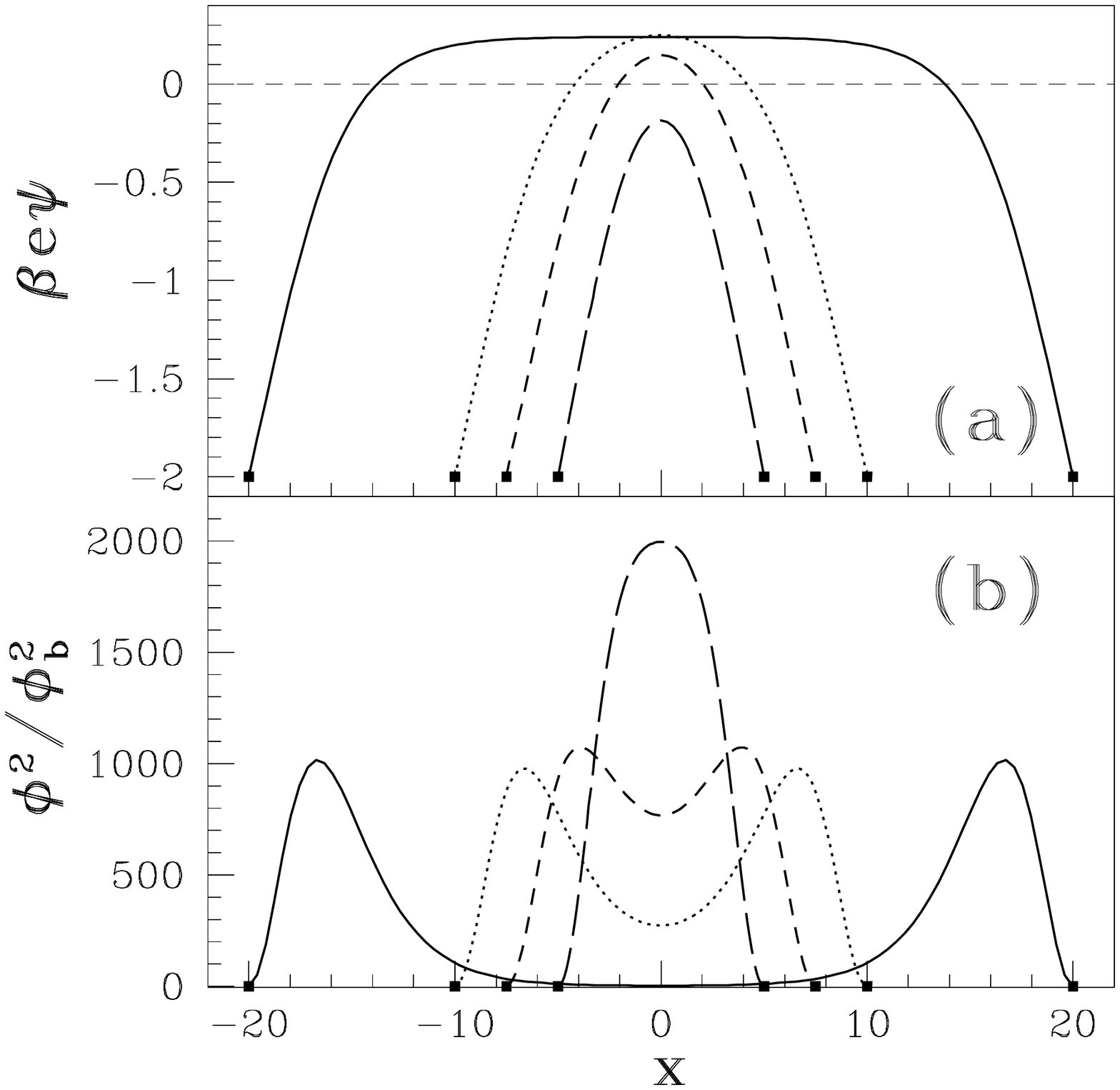} }}
\end{figure}
\vfill

\begin{figure}[tbh]
  {\Large Fig.~4}
  \bigskip\bigskip\bigskip

  \epsfxsize=0.5\linewidth
  \centerline{\hbox{ \epsffile{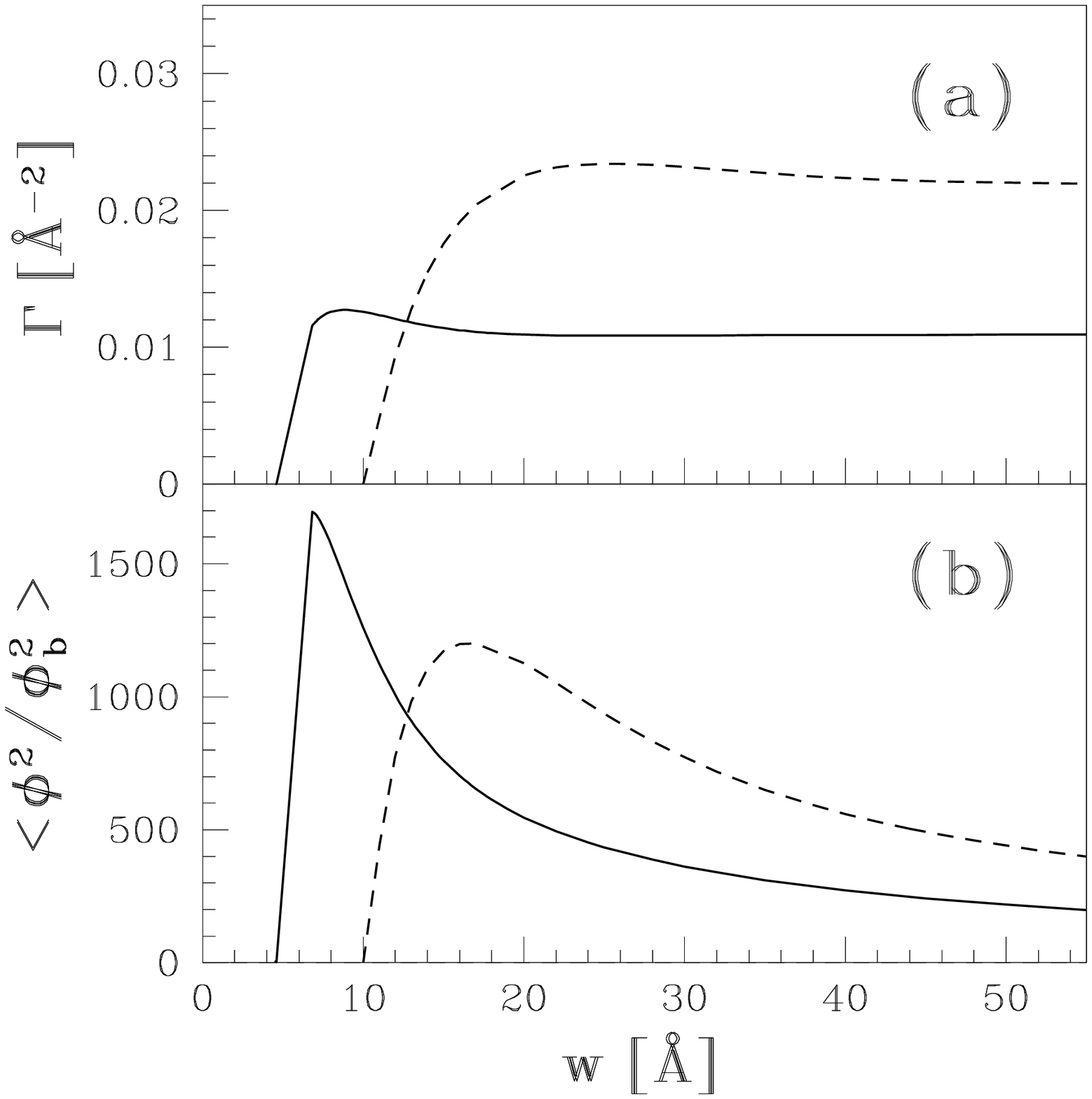} }}
\end{figure}
\vfill

\begin{figure}[tbh]
  {\Large Fig.~5}
  \bigskip\bigskip\bigskip

  \epsfxsize=0.5\linewidth
  \centerline{\hbox{ \epsffile{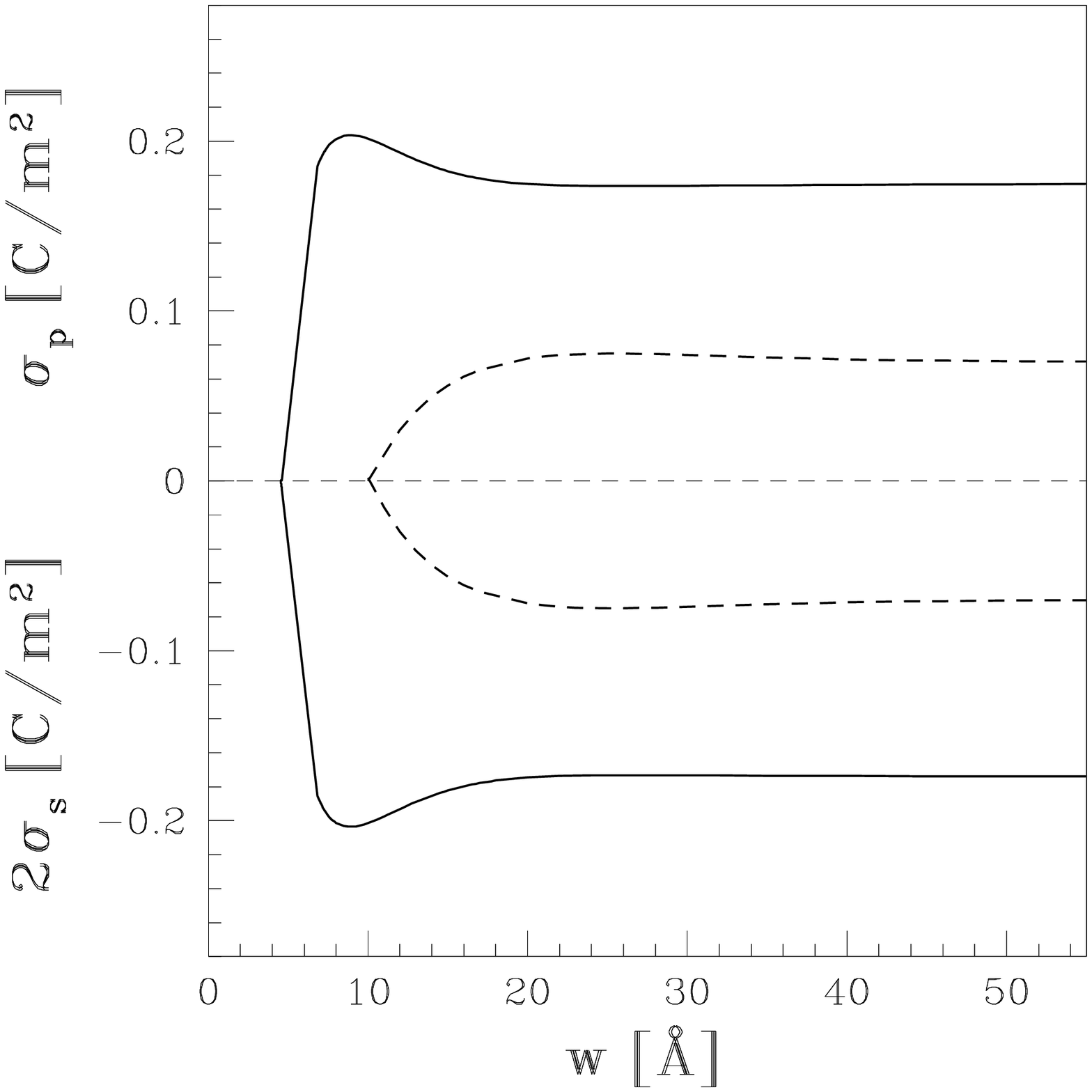} }}
\end{figure}
\vfill

\begin{figure}[tbh]
  {\Large Fig.~6}
  \bigskip\bigskip\bigskip

  \epsfxsize=0.5\linewidth
  \centerline{\hbox{ \epsffile{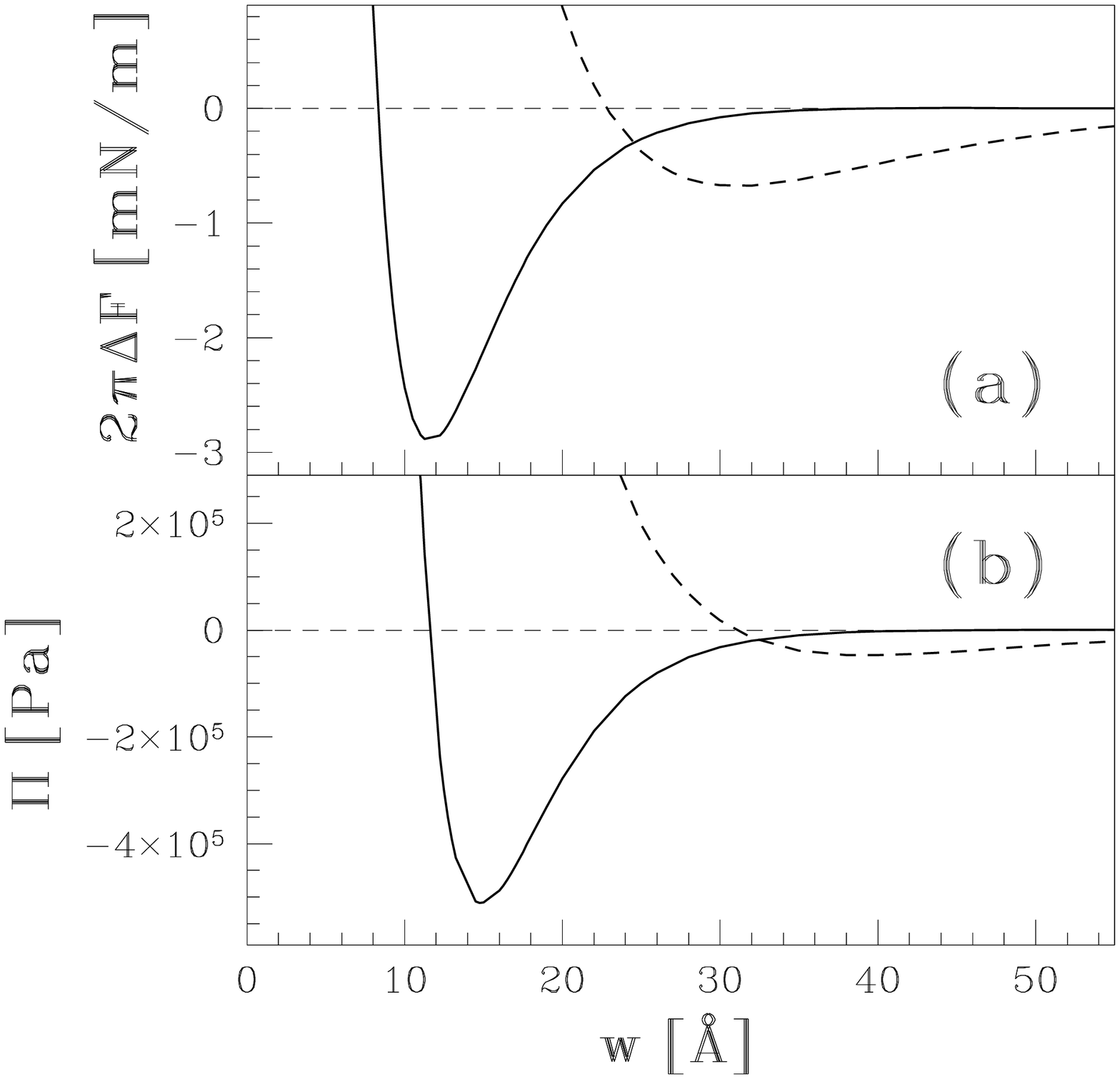} }}
\end{figure}
\vfill

\begin{figure}[tbh]
  {\Large Fig.~7}
  \bigskip\bigskip\bigskip
  
  \epsfxsize=0.5\linewidth
  \centerline{\hbox{ \epsffile{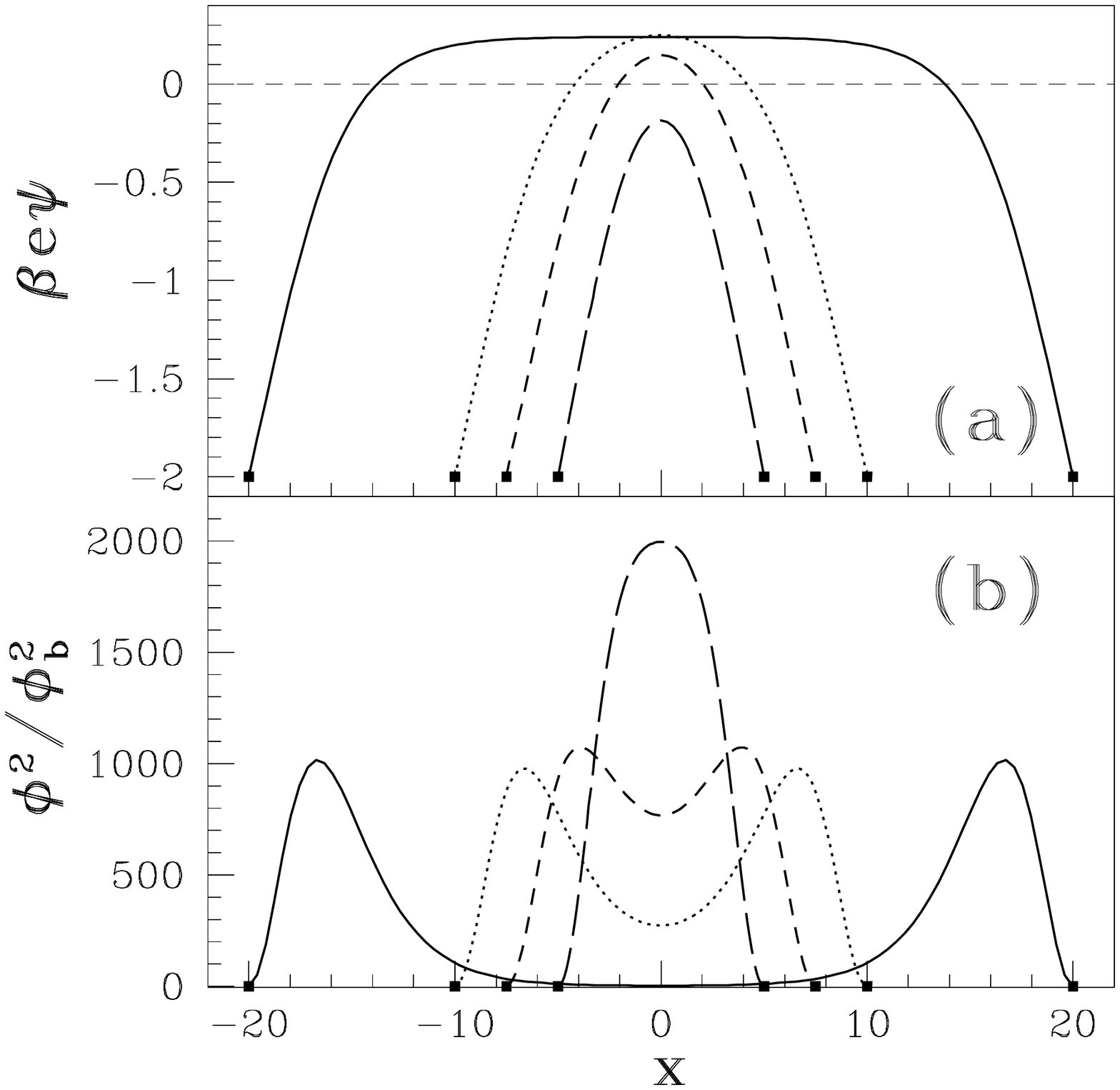} }}
\end{figure}
\vfill

\begin{figure}[tbh]
  {\Large Fig.~8}
  \bigskip\bigskip\bigskip

  \epsfxsize=0.5\linewidth
  \centerline{\hbox{ \epsffile{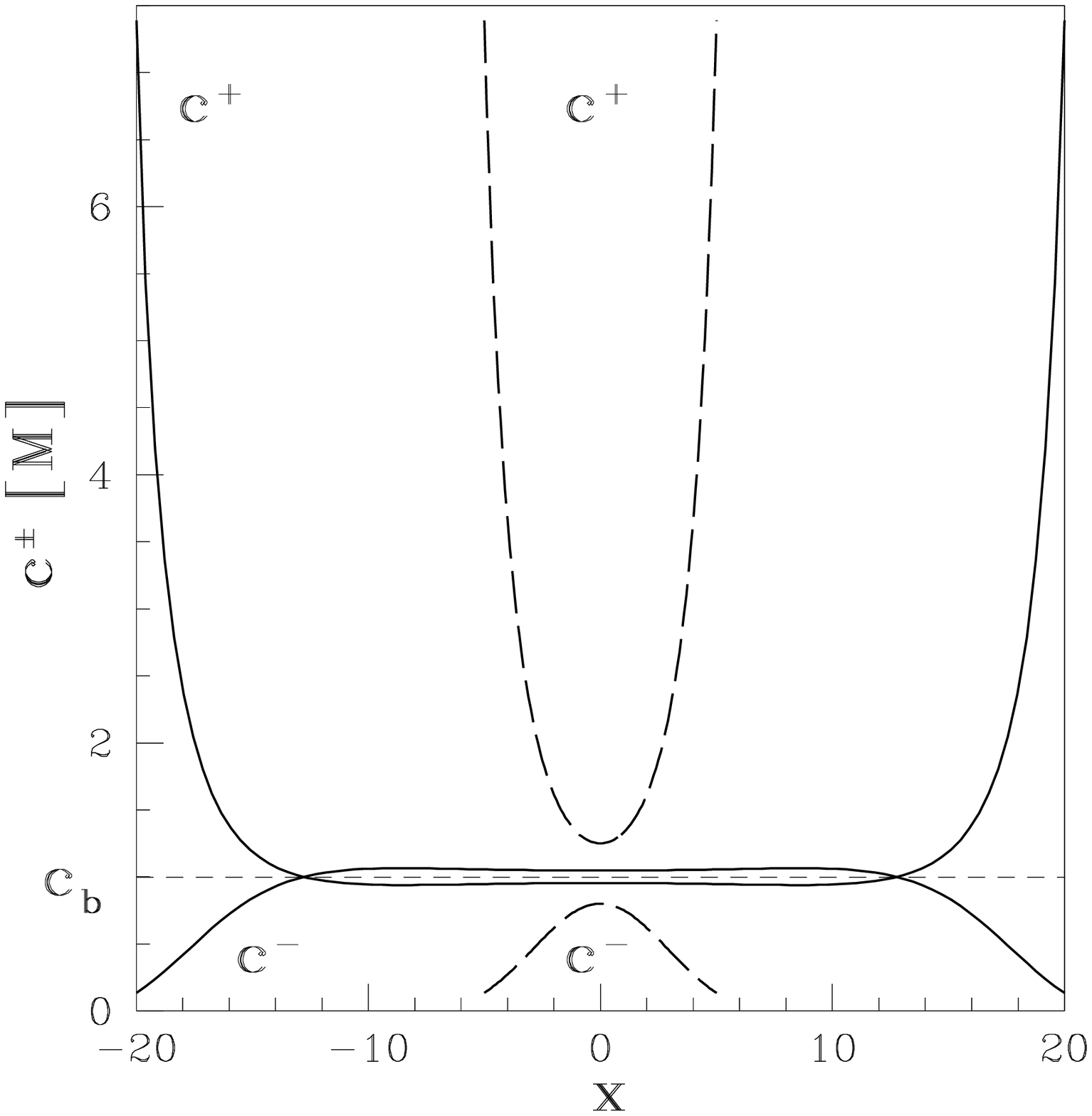} }}
\end{figure}
\vfill

\begin{figure}[tbh]
  {\Large Fig.~9}
  \bigskip\bigskip\bigskip

  \epsfxsize=0.5\linewidth
  \centerline{\hbox{ \epsffile{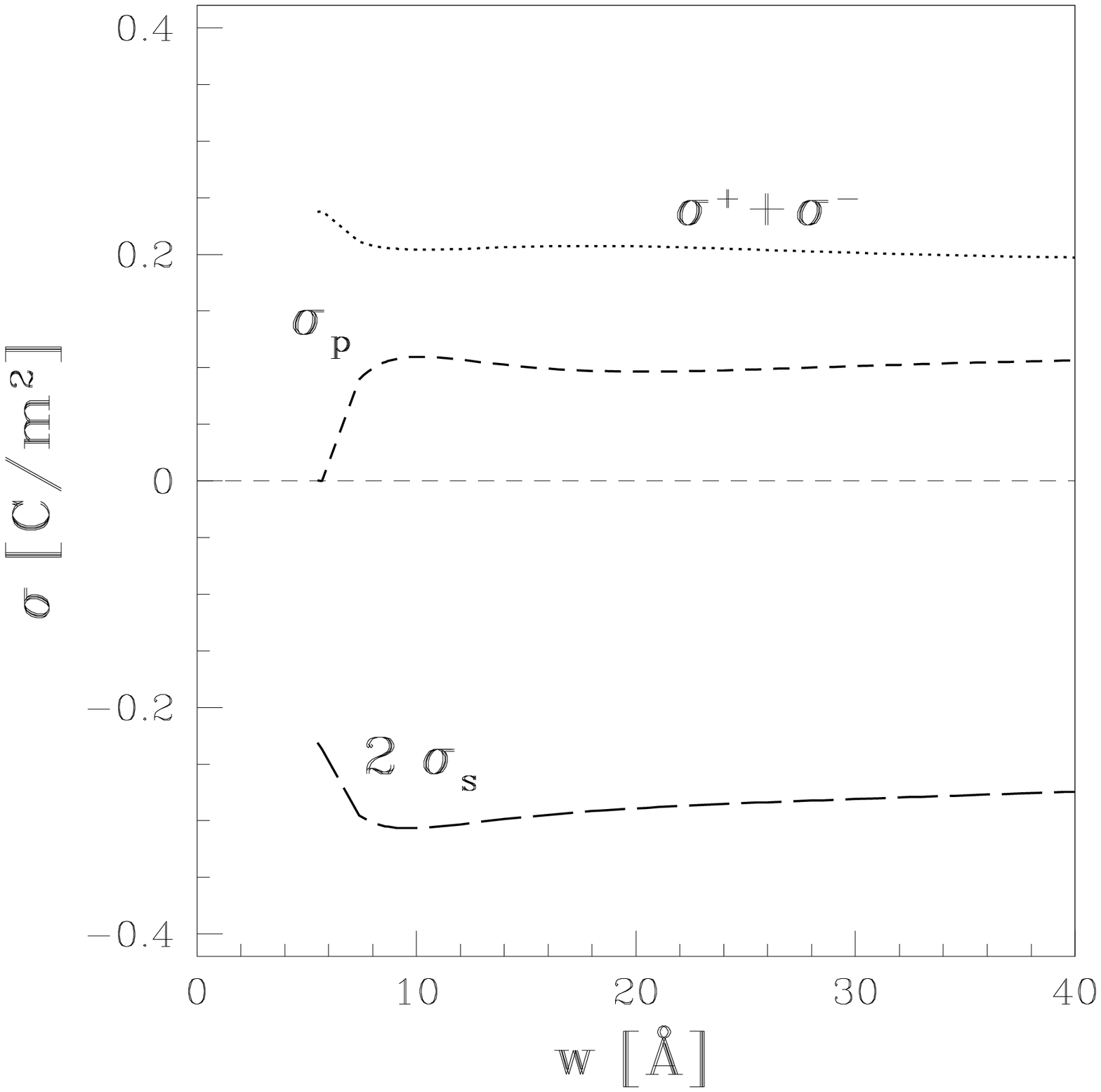} }}
\end{figure}
\vfill

\begin{figure}[tbh]
  {\Large Fig.~10}
  \bigskip\bigskip\bigskip

  \epsfxsize=0.5\linewidth
  \centerline{\hbox{ \epsffile{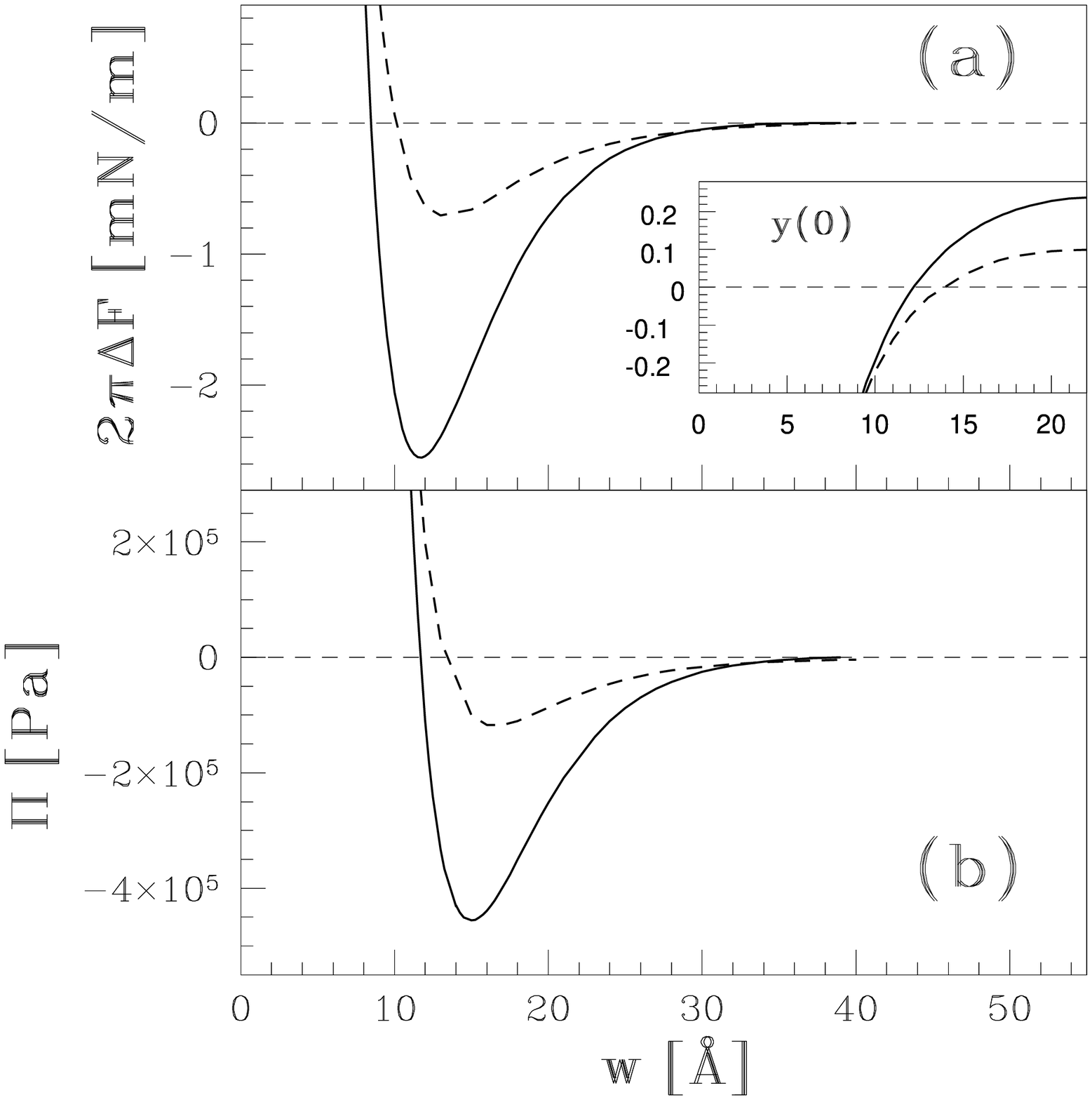} }}
\end{figure}
\vfill

\begin{figure}[tbh]
  {\Large Fig.~11}
  \bigskip\bigskip\bigskip

  \epsfxsize=0.5\linewidth
  \centerline{\hbox{ \epsffile{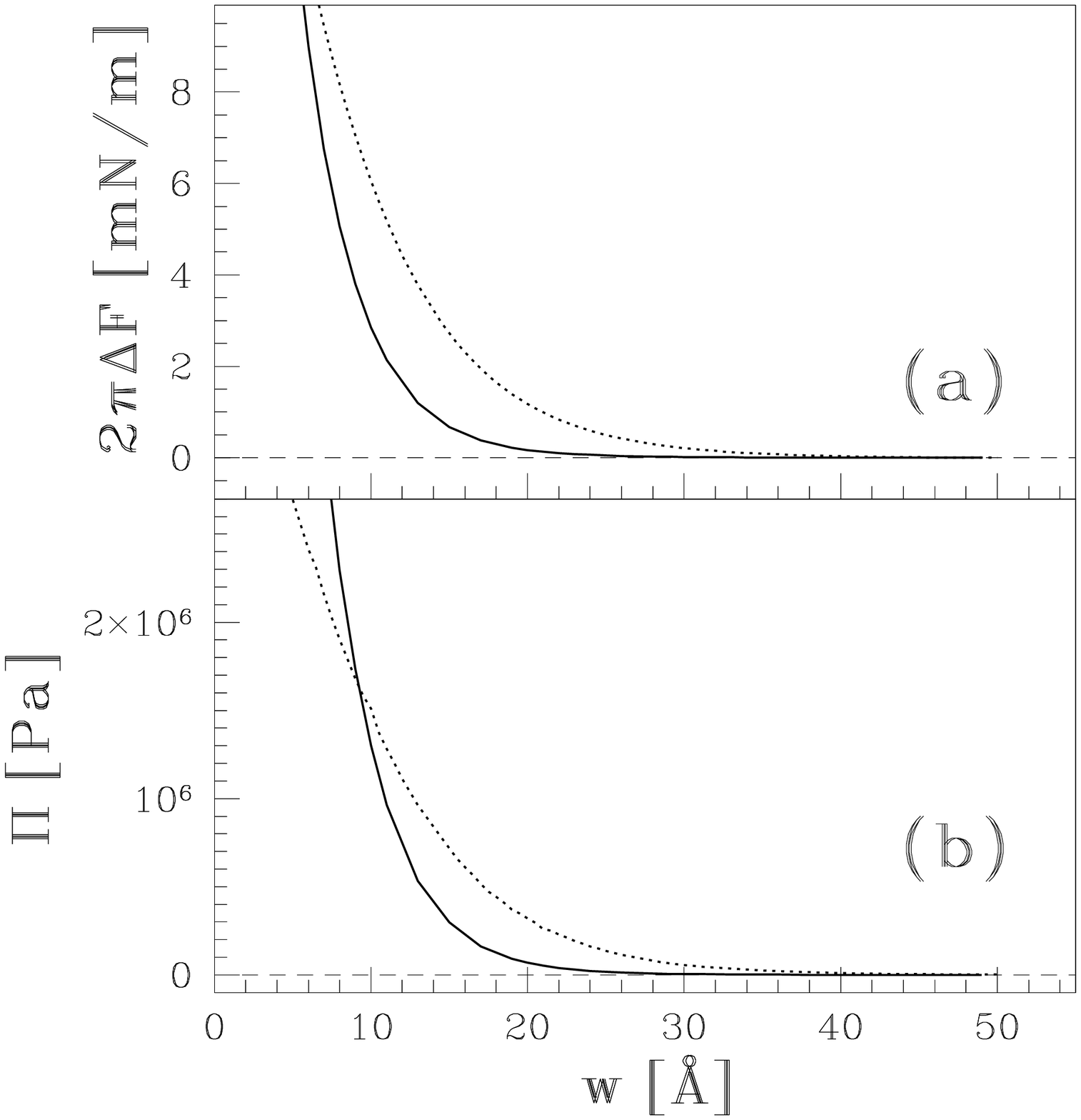} }}
\end{figure}
\vfill

\begin{figure}[tbh]
  {\Large Fig.~12}
  \bigskip\bigskip\bigskip

  \epsfxsize=0.5\linewidth
  \centerline{\hbox{ \epsffile{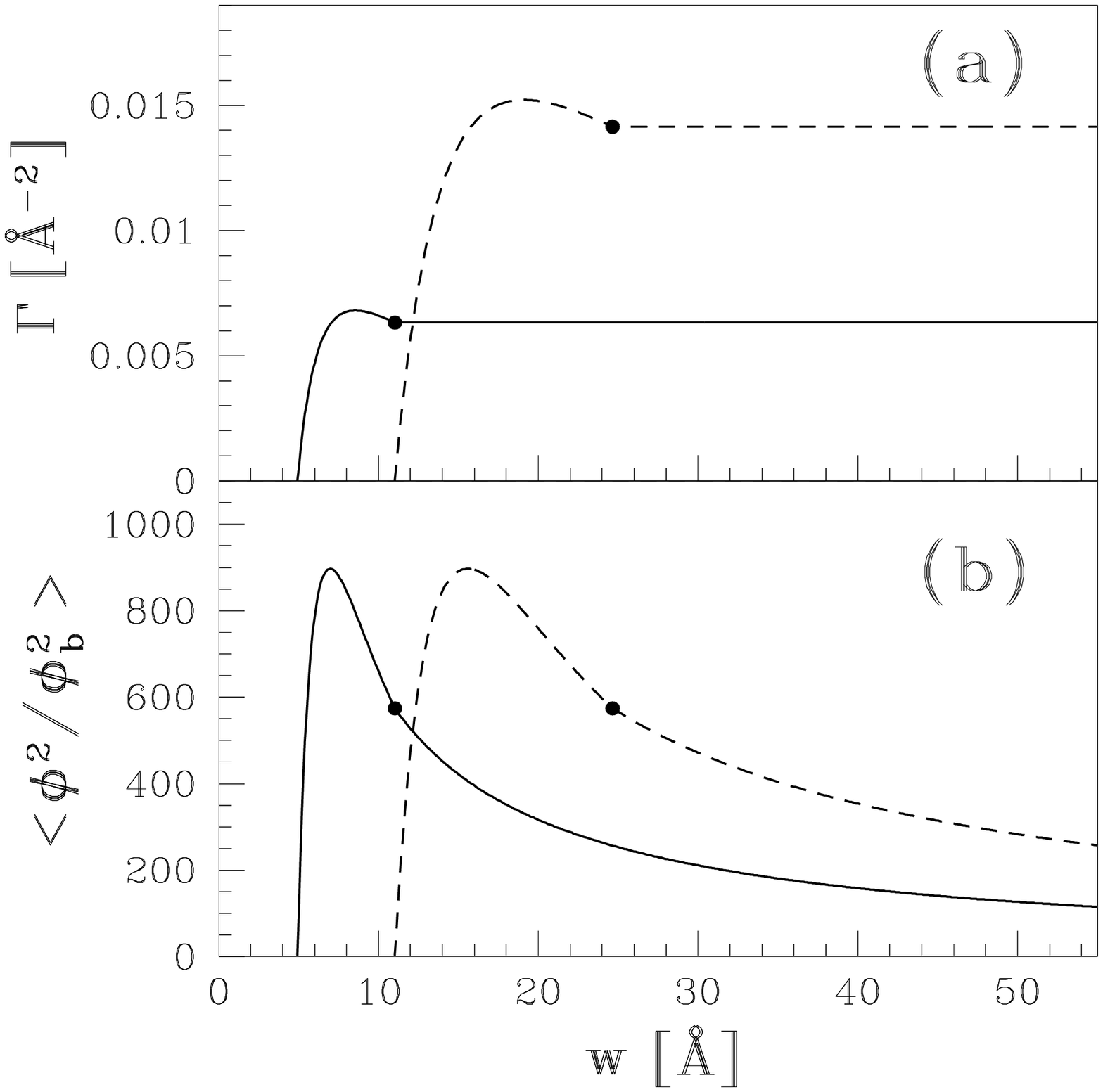} }}
\end{figure}
\vfill

\begin{figure}[tbh]
  {\Large Fig.~13}
  \bigskip\bigskip\bigskip

  \epsfxsize=0.5\linewidth
  \centerline{\hbox{ \epsffile{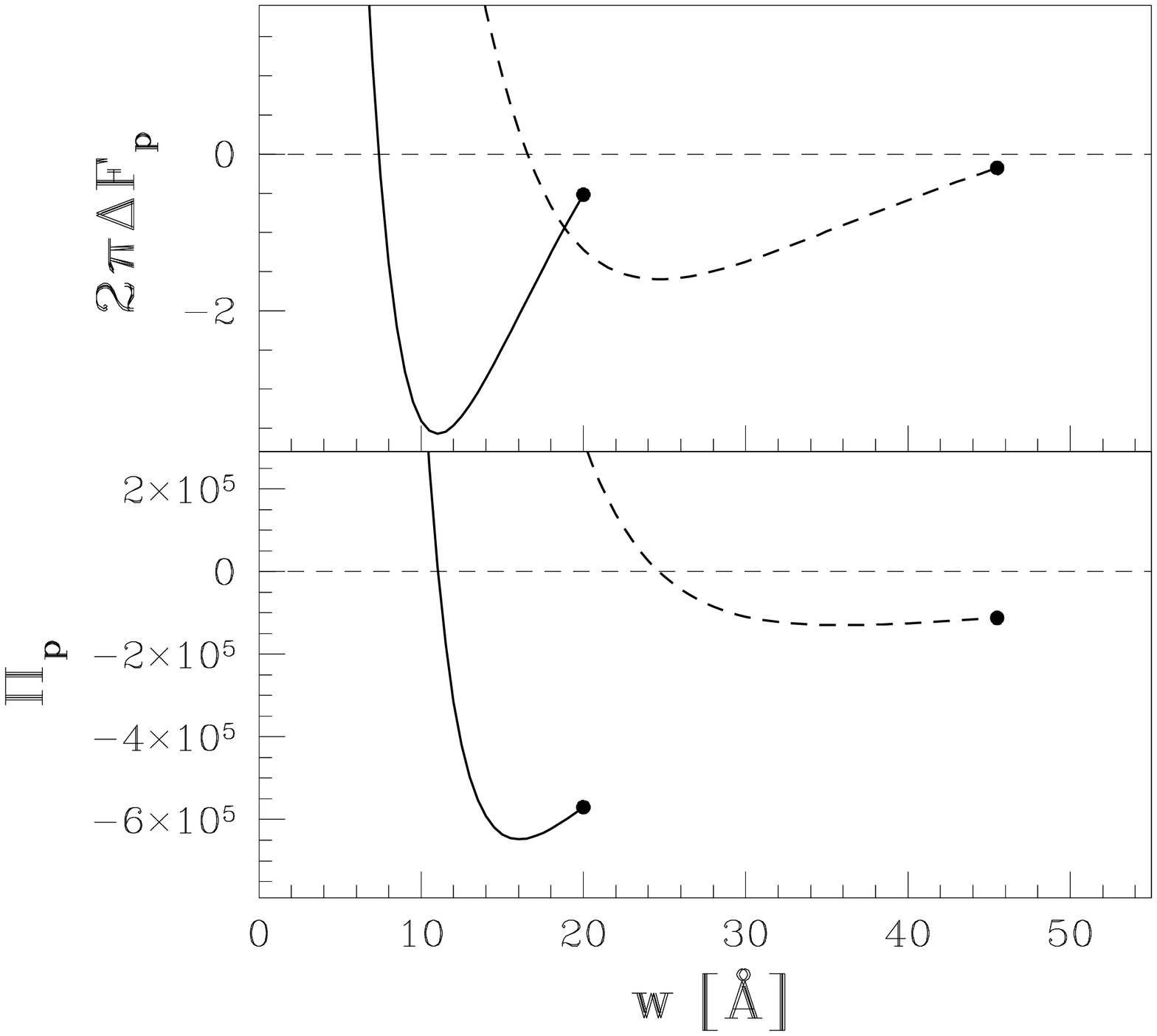} }}
\end{figure}
\vfill

\begin{figure}[tbh]
  {\Large Fig.~14}
  \bigskip\bigskip\bigskip

  \epsfxsize=0.5\linewidth
  \centerline{\hbox{ \epsffile{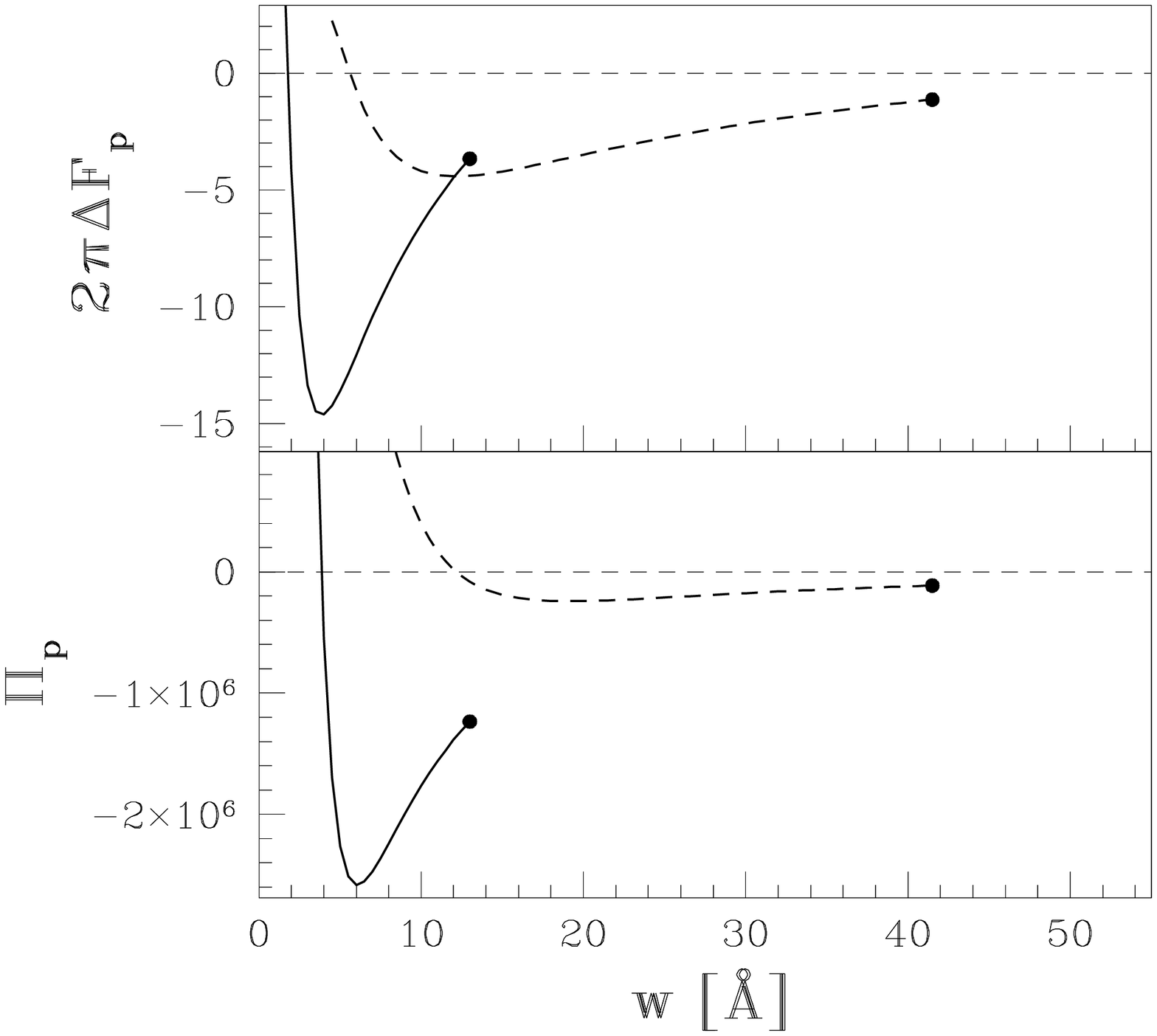} }}
\end{figure}
\vfill

\begin{figure}[tbh]
  {\Large Fig.~15}
  \bigskip\bigskip\bigskip

  \epsfxsize=0.5\linewidth
  \centerline{\hbox{ \epsffile{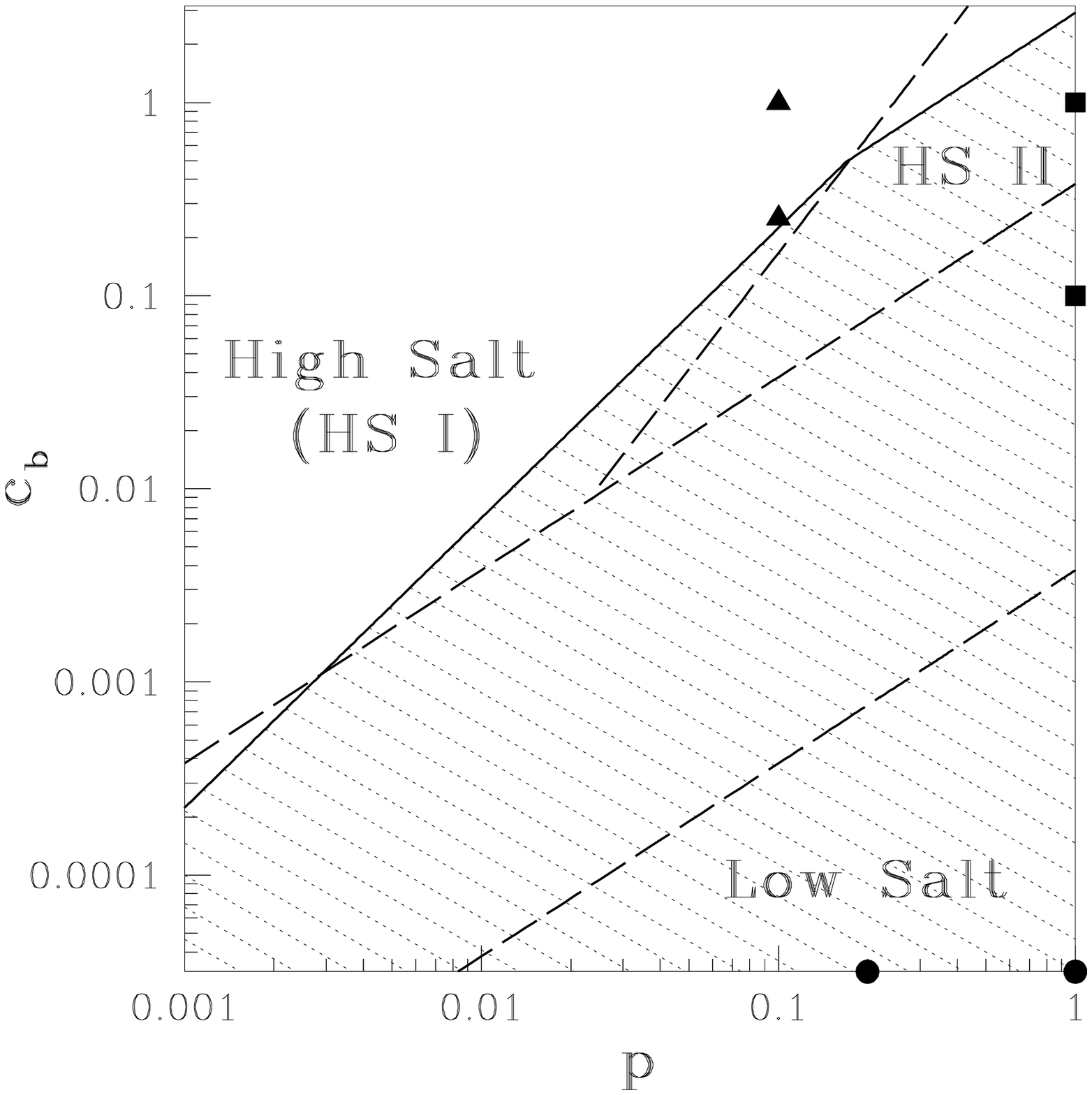} }}
\end{figure}
\vfill

\begin{figure}[tbh]
  {\Large Fig.~16}
  \bigskip\bigskip\bigskip

  \epsfxsize=0.5\linewidth
  \centerline{\hbox{ \epsffile{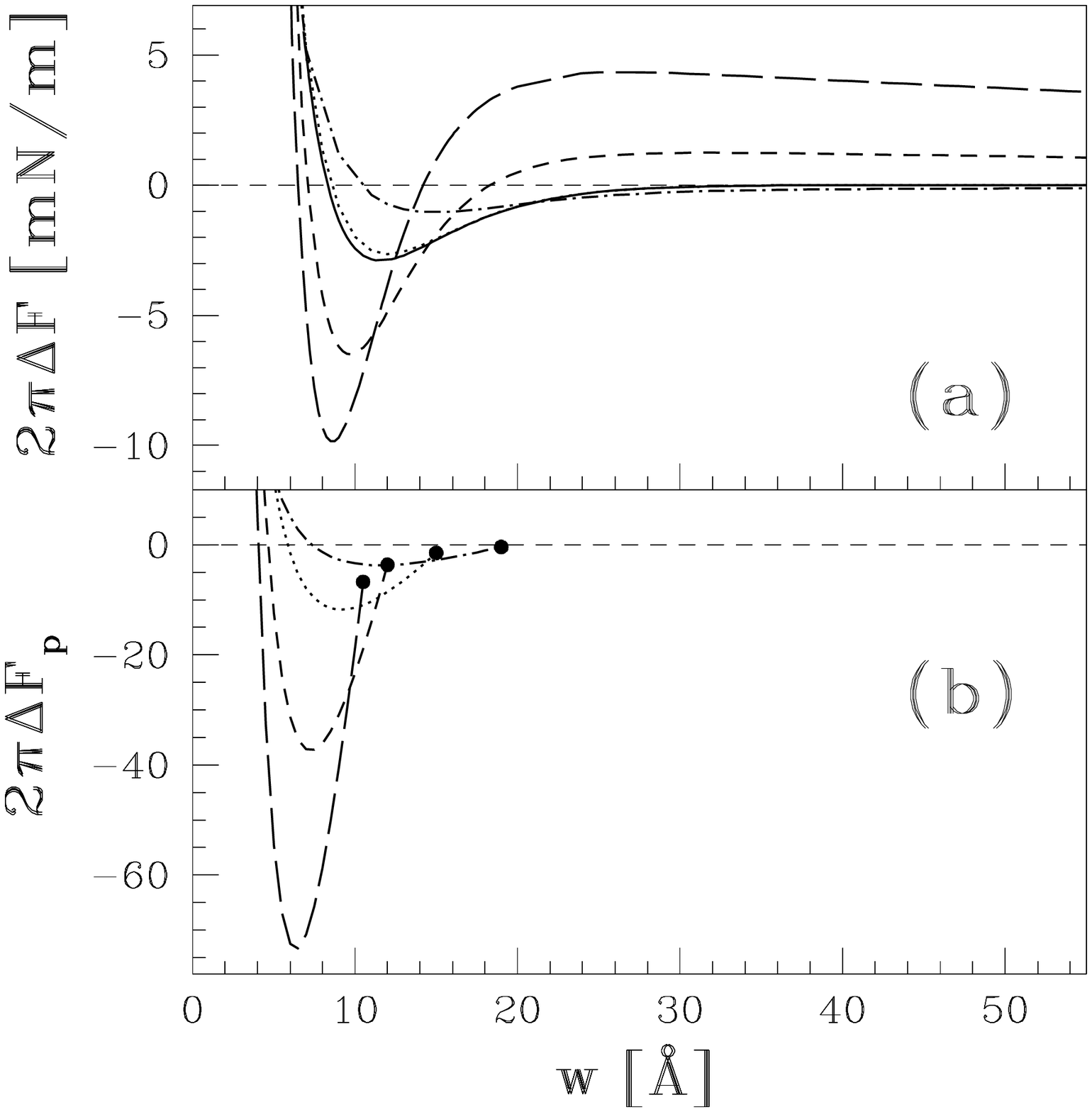} }}
\end{figure}
\vfill

%-----------------------------------------------------------

\begin{thebibliography}{99}
\bibitem{beyondPB}
    Recent measurements of inter--surface forces in the 
presence of asymmetric electrolytes (such as CaCl$_2$)
and without polyelectrolytes
followed by theoretical calculations using the 
hyper-netted chain (HNC) scheme have indicated that 
ion correlations beyond mean field  can lead to attraction
of symmetric surfaces
at very short separations of a few Angstroms, e.g., 
ref.~\cite{Kjellander}.
\bibitem{depletion}
  Depletion forces contribute to inter--surface attraction.
However this contribution is proportional to $k_BT v\phi_b^4$
and is much smaller than the electrostatic repulsion.
\bibitem{wmin}
Since the linear profile $h_1(z)$ is a better approximation at small
distances we use it in order to estimate the lower
cutoff $w_{\rm min}$, while the coefficients of
the parabolic profile $h_2(z)$ is used in the calculation of 
the single surface adsorption length $D_1$.
\end{thebibliography}
\end{document}